\documentclass[preprint,aps,amsmath,amssymb,nofootinbib,12pt]{revtex4}

\usepackage{epsfig}
\usepackage{slashed}
\usepackage{graphicx}
\usepackage{multirow,color}
\usepackage{amsmath}
\usepackage{float}
\usepackage{diagbox}
\usepackage{CJK}
\usepackage{color}
\usepackage{xcolor}
\usepackage{times}
\usepackage{subfigure}
\usepackage{bm}
\usepackage{braket}
\usepackage{booktabs}
\usepackage{array}
\usepackage[mathscr]{euscript}
\usepackage{caption}
\usepackage{makecell}
\usepackage{epstopdf}
\usepackage{hyperref}
\usepackage{amsmath,amssymb,amsthm,amsxtra,overpic,bbm,bm,epsfig,ulem,color,multirow}

\makeatletter

\newcommand{\Rmnum}[1]{\expandafter\@slowromancap\romannumeral #1@}
\makeatother
\graphicspath{{fig/}}
\begin{document}
\title{Prospects for probing neutral vector-like leptons via pair production at a muon collider}
\author{Chong-Xing Yue$^{1,2}$}
\thanks{cxyue@lnnu.edu.cn}
\author{Mei-Shu-Yu Wang$^{1,2}$}
\thanks{wmsy0108@163.com~(Corresponding author)}
\author{Xin-Yang Li$^{1,2}$}
\thanks{lxy91108@163.com}
\author{Si-Yu Zhang$^{1,2}$}
\thanks{m16630827711@163.com}

\affiliation{
$^1$Department of Physics, Liaoning Normal University, Dalian 116029, China\\
$^2$Center for Theoretical and Experimental High Energy Physics, Liaoning Normal University, Dalian 116029, China
}

\begin{abstract}
 Vector-like leptons (VLLs) are well-motivated candidates for physics beyond the standard model.
 We investigate the sensitivity of the $6$ TeV muon collider to the neutral doublet VLL (denoted by $N$) via its pair production in an SU(2)$_L$ doublet model with hypercharge $Y=-\frac{1}{2}$. Assuming mixing with muons, we study the subsequent decay $N\rightarrow W^+ \mu^-$ and analyze two representative signals, namely $4j2\mu$ and $2\ell2\mu E\mkern-10.5 mu/$, arising from the hadronic and leptonic decays of the $W$ boson, respectively. The signal and background events are simulated within a Monte Carlo framework, and a cut-based analysis is performed at the $6$ TeV muon collider with an integrated luminosity $\mathcal{L}=4$ ab$^{-1}$ for the unpolarized and polarized beam configurations, $(P_{\mu^+},P_{\mu^-})=(0,0)$ and $(P_{\mu^+},P_{\mu^-})=(-1,1)$.
 We consider VLL masses in the range of $1300-2980$ GeV and evaluate the corresponding search sensitivity.
 Our results show that the $6$ TeV muon collider could provide strong sensitivity to neutral doublet VLLs through  the $4j2\mu$ and $2\ell2\mu E\mkern-10.5 mu/$ signals, particularly for the idealized fully polarized beam configuration, under the detector, polarization, luminosity and background assumptions adopted in this analysis.
\end{abstract}
\maketitle

%%%%%%%%%%%%%%%%%%%%%%%%%%%%%%%%%%%%%%%%%%%%%%%%%%
%%%%%%%%%%%%%%%%%%%%%%%%%%%%%%%%%%%%%%%%%%%%%%%%%%
\section{Introduction}
%%%%%%%%%%%%%%%%%%%%%%%%%%%%%%%%%%%%%%%%%%%%%%%%%%
%%%%%%%%%%%%%%%%%%%%%%%%%%%%%%%%%%%%%%%%%%%%%%%%%%
The Standard Model (SM) of particle physics~\cite{Salam:1968rm,Weinberg:1967tq,Glashow:1961tr} provides a well-established framework for describing the fundamental particles and their interactions. Based on the gauge symmetry group SU(3)$_C \times$ SU(2)$_L \times$ U(1)$_Y$, it successfully accounts for a wide range of experimental observations across different energy scales. Nevertheless, the SM does not address several important open questions, including the gauge hierarchy problem~\cite{Feng:2013pwa}, the nature of dark matter~\cite{Duffy:2009ig,Chadha-Day:2021szb} and the origin of neutrino masses~\cite{Gonzalez-Garcia:2007dlo}. These limitations indicate that the SM is likely not a complete description of particle physics. Theories beyond the Standard Model (BSM) have been proposed to address these issues and many of them predict the existence of new particles.
Among these, vector-like leptons (VLLs) are possible candidates for BSM physics. They have been studied in relation to several anomalies, including the electron and muon $g-2$ discrepancies~\cite{Hiller:2019mou,Frank:2020smf,Dermisek:2021ajd, Brune:2022rlo,Guedes:2022cfy,Mohling:2024qvk,Daberstiel:2026wxe}, the difference between the observed and predicted $W$-boson mass~\cite{He:2022zjz,Kawamura:2022fhm} and the Cabibbo angle anomaly~\cite{Belfatto:2019swo,Grossman:2019bzp,Crivellin:2020ebi}. They may also provide explanations for open issues such as dark matter, electroweak vacuum stability and Higgs sector dynamics. In addition, VLLs have been studied within the flavour democracy hypothesis as a possible way to address the fermion mass hierarchy~\cite{Dagli:2026lus}. As a result, VLLs arise naturally in many theoretical frameworks, including supersymmetric models~\cite{Martin:2009bg,Zheng:2019kqu,Kong:2010qd,Endo:2011mc,Araz:2018uyi}, composite Higgs models~\cite{Anastasiou:2009rv,Gillioz:2012se}, extra-dimensional scenarios~\cite{Agashe:2006wa,Huang:2012kz}, left-right symmetric models~\cite{Tsai:1972sg,Mohapatra:1974gc,Senjanovic:1975rk,Mohapatra:1977mj} and grand unified theories~\cite{Nevzorov:2012hs,Dorsner:2014wva,Joglekar:2016yap}.

Unlike the chiral leptons in the SM, VLLs have left- and right-handed components that transform identically under the gauge symmetries. This allows gauge-invariant mass terms to be introduced independently of the Higgs mechanism. In contrast, extra chiral leptons must acquire their masses through Yukawa interactions, which can induce non-decoupling effects and lead to sizable corrections to precision observables. As a result, extra chiral leptons are tightly constrained by precision measurements~\cite{CMS:2024bni,delAguila:2008pw,Ishiwata:2013gma,Thomas:1998wy}. Consequently, VLLs are generally less constrained and exhibit a wider range of collider phenomenology. Depending on whether they mix with first-, second- or third-generation SM leptons $(\ell=e,\mu,\tau)$, they are commonly referred to as vector-like electrons, muons or $\tau$-leptons, respectively.

Early studies at the Large Electron-Positron (LEP) experiments have excluded VLLs with masses below $101.2$ GeV at the 95$\%$ confidence level (C.L.), independently of their SU(2)$_L$ representation~\cite{L3:2001xsz}. Both the ATLAS and CMS Collaborations have searched for VLLs at the Large Hadron Collider (LHC). The CMS Collaboration has excluded vector-like $\tau$-leptons with masses below $1045$ GeV in the doublet scenario and in the range of $125-150$ GeV in the singlet scenario based on data with a center-of-mass energy $\sqrt{s} = 13$ TeV and an integrated luminosity  $\mathcal{L}=$ $138$ fb$^{-1}$~\cite{CMS:2022nty}.
More recent results from the ATLAS Collaboration, obtained with $\sqrt{s} = 13$ TeV and $\mathcal{L}=$ $140$ fb$^{-1}$, exclude vector-like electrons with masses below $1220$ GeV in the doublet scenario and $320$ GeV in the singlet scenario, vector-like muons below $1270$ GeV (doublet) and $400$ GeV (singlet)~\cite{ATLAS:2024mrr} and $\tau$-leptons in the mass range of $130-900$ GeV in the doublet scenario~\cite{ATLAS:2023sbu}.

Beyond these experimental constraints, numerous theoretical and phenomenological studies have explored the properties and signatures of VLLs at collider experiments. These studies include investigations at electron-positron~\cite{Bahrami:2016has,Mahmoud:2024sby,Yue:2024ftz,Yue:2024sds,Cao:2023smj, Liu:2025ori,Yue:2025xvk,Bhattacharya:2021ltd} and hadron colliders~\cite{Graham:2009gy,Bernreuther:2023uxh,Endo:2011xq,Ellis:2014dza,Kumar:2015tna,Xu:2018pnq,Dermisek:2015oja,Dermisek:2014qca, Dermisek:2013gta, Freitas:2020ttd,Frampton:1999xi,Falkowski:2013jya,Guedes:2021oqx,Bhattiprolu:2019vdu,Sher:1995tc,Chala:2020odv}, carried out within both specific models and model-independent frameworks. Taken together, they indicate that the search for VLLs remains an active area of research, with both prompt and long-lived signatures studied across a wide range of scenarios. In addition to studies at electron-positron and hadron colliders, several works have investigated the potential of the muon collider for probing VLLs~\cite{Guo:2023jkz,Morais:2021ead,Ghosh:2023xbj}.

The muon collider has recently emerged as a promising option for a next-generation high-energy collider. By colliding muons instead of electrons or protons, it combines the advantages of both electron-positron and hadron colliders~\cite{Greco:2016izi,Delahaye:2019omf,Long:2020wfp,MuonCollider:2022xlm,Accettura:2023ked, InternationalMuonCollider:2024jyv,Black:2022cth}. Owing to the much larger mass of the muon compared to the electron, synchrotron radiation is strongly suppressed, allowing muons to be accelerated to multi-TeV energies in a relatively compact circular collider. In contrast to proton-proton collisions, muon-muon collisions provide a well-defined initial state and a comparatively clean hard-scattering environment, although beam-induced backgrounds from muon decays remain an important detector challenge. For these reasons, the muon collider is a promising option for studying high-energy processes and searching for new physics at the multi-TeV scale. Current muon collider studies commonly consider center-of-mass energies of around $3$ TeV and $10$ TeV as representative reference stages, with integrated luminosities of $1$ ab$^{-1}$ and $10$ ab$^{-1}$, respectively~\cite{Black:2022cth,Han:2020uid,Costantini:2020stv}.
An intermediate setup with $\sqrt{s}=6$ TeV and an integrated luminosity of $4$ ab$^{-1}$ has also been considered in several phenomenological studies~\cite{Huang:2021biu,Bandyopadhyay:2024plc,Yang:2021zak,Han:2020pif}. We adopt this setup in the present work as an intermediate phenomenological benchmark, rather than as an official machine stage.

Given their high center-of-mass energy and well-defined initial state, muon colliders provide attractive opportunities to search for heavy VLLs. In this work, we consider an SU(2)$_L$ doublet VLL model in which the VLLs mix with the second-generation SM leptons. The doublet contains the neutral state $N$ and its approximately mass-degenerate charged partner $E$. Since charged VLL production at muon colliders has already been studied in Refs.~\cite{Guo:2023jkz,Morais:2021ead, Ghosh:2023xbj}, we focus here on the neutral $N\bar{N}$ pair-production channel. In the small-mixing limit of the model, the $N\bar{N}$ pair-production cross section depends on the VLL representation and mass, but not on the flavour choice or decay branching ratios. The flavour choice determines the decay channels and the corresponding experimental signatures.
Motivated by these considerations, we investigate $\mu^+\mu^-\rightarrow N\bar{N}$ for masses ranging from $1300$ to $2980$ GeV at the $6$ TeV muon collider. Both unpolarized and polarized beam configurations are considered to examine the impact of beam polarization on the search sensitivity.

The structure of this paper is as follows. In Section II, we introduce the doublet VLL model and present the relevant interactions between the VLLs and the SM particles. We also discuss the theoretical constraints on the model parameters and specify the benchmark setup adopted in our analysis. In Section III, we investigate the prospects for probing neutral VLLs through pair production at the $6$ TeV muon collider.
Detector simulations are performed for the selected signals, and the corresponding signal significances are evaluated. Finally, we summarize our conclusions in Section IV.

%%%%%%%%%%%%%%%%%%%%%%%%%%%%%%%%%%%%%%%%%%%%%%%%%%
%%%%%%%%%%%%%%%%%%%%%%%%%%%%%%%%%%%%%%%%%%%%%%%%%%
\section{The theory framework}
%%%%%%%%%%%%%%%%%%%%%%%%%%%%%%%%%%%%%%%%%%%%%%%%%%
%%%%%%%%%%%%%%%%%%%%%%%%%%%%%%%%%%%%%%%%%%%%%%%%%%

The VLL model considered here is based on the SU(2)$_L$ doublet VLL model introduced in Ref.~\cite{Kumar:2015tna}. In the original framework, the VLLs predominantly mix with the third-generation SM leptons,
whereas in this work we consider the vector-like muon scenario as a benchmark, since the decay $N\rightarrow W^+ \mu^-$ provides a clean benchmark final state for the collider analysis. The new fermionic doublet is denoted by
\begin{eqnarray}
\begin{split}
\label{eq:2.1}
L'=
\begin{pmatrix}
N,\
E
\end{pmatrix}.
\end{split}
\end{eqnarray}
In the two-component notation used in Ref.~\cite{Kumar:2015tna}, it transforms under the SM gauge group SU(3)$_C \times$ SU(2)$_L \times$ U(1)$_Y$ as
\begin{eqnarray}
\begin{split}
\label{eq:2.2}
L' \sim (1,2,-\tfrac{1}{2}).
\end{split}
\end{eqnarray}
Here, $N$ and $E$ denote a neutral and charged VLL, respectively. Since the left- and right-handed components of the VLL doublet transform identically under the SM gauge group, a gauge-invariant Dirac mass term can be introduced independently of electroweak symmetry breaking.

In the doublet VLL model for the second generation, mixing between the VLLs and the SM muon sector is induced through Yukawa interactions involving the Higgs field. The masses of the VLLs are primarily determined by a common gauge-invariant vector-like mass parameter $M_{L}$. The relevant Lagrangian can be written as
\begin{eqnarray}
\begin{split}
\label{eq:2.3}
-\mathcal{L}
	=M_{L} L'\bar{L}'+\epsilon H L'\bar{\mu}+y_\mu H L_\mu\bar{\mu}+\mathrm{c.c.},
\end{split}
\end{eqnarray}
where $L'$ denotes the vector-like lepton doublet, $H$ represents the SM Higgs scalar doublet, $L_\mu=(\nu_\mu,\mu)_L$ denotes the SM second-generation
lepton doublet in the gauge eigenstate basis, $y_\mu$ is the SM $\mu$ Yukawa coupling, and $\epsilon$ is the mixing Yukawa coupling.

After electroweak symmetry breaking, the Higgs field acquires a vacuum expectation value $v\approx 174$ GeV, generating mass terms for both the muon and the vector-like charged lepton. In the $(\mu,E)$ basis, the corresponding mass matrix can be written as
\begin{eqnarray}
\begin{split}
\label{eq:2.4}
\mathcal{M}
=
\begin{pmatrix}
y_\mu v & 0\\
\epsilon v & M_{L}
\end{pmatrix}.
\end{split}
\end{eqnarray}
The physical mass eigenstates are obtained by diagonalizing the above mass matrix. Throughout this work, we focus on the parameter region satisfying $\epsilon v\ll M_{L}$, where the mixing between the muon and the VLL is small. In this limit, the muon mass remains approximately equal to its SM value, while both vector-like states have masses close to the common mass parameter $M_{L}$,
\begin{eqnarray}
\begin{split}
\label{eq:2.5}
m_\mu \simeq y_\mu v,\quad M_E\simeq M_{L},\quad M_N\simeq M_{L}.
\end{split}
\end{eqnarray}
Since the neutral and charged VLLs belong to the same SU(2)$_L$ doublet and are governed by the same vector-like mass parameter, their masses are degenerate at tree level. Consequently, we adopt $M_N\simeq M_E\simeq M_L$ throughout this work. Although electroweak radiative corrections induce a small mass splitting between the neutral and charged states, the resulting difference is typically only of the order of a few hundred MeV. This effect is therefore negligible compared with the TeV-scale masses and is neglected in the present analysis.

We now turn to the gauge interactions of the VLLs in the doublet VLL model. Neglecting the small mixing between the VLLs and the second-generation SM leptons, the relevant interaction Lagrangian for $E$ and $N$ is given by~\cite{Kumar:2015tna}
\begin{eqnarray}
\begin{split}
\label{eq:2.6}
\mathcal{L}_{\rm int} =&\frac{e}{\sqrt{2}s_W} W_\mu^{+}(\bar{E}^{\dagger}\bar{\sigma}^\mu \bar{N} + N^{\dagger} \bar{\sigma}^\mu E) + \frac{e}{\sqrt{2}s_W} W_\mu^{-}(E^{\dagger}\bar{\sigma}^\mu N + \bar{N}^{\dagger} \bar{\sigma}^\mu \bar{E}) - e A_\mu (E^{\dagger}\bar{\sigma}^\mu E - \bar{E}^{\dagger} \bar{\sigma}^\mu \bar{E})\\[2mm]
&+
\frac{e}{c_W s_W}(s_W^2-\frac{1}{2})Z_\mu (E^{\dagger}\bar{\sigma}^\mu E - \bar{E}^{\dagger} \bar{\sigma}^\mu \bar{E}) + \frac{e}{2c_W s_W}Z_\mu (N^{\dagger}\bar{\sigma}^\mu N - \bar{N}^{\dagger} \bar{\sigma}^\mu \bar{N}),
\end{split}
\end{eqnarray}
where $e$ is the electromagnetic coupling, and $s_W=\sin\theta_W$ and $c_W=\cos\theta_W$ represent the sine and cosine of the weak mixing angle, respectively. Here, $A_\mu, Z_\mu$ and $W_\mu^\pm$ denote the photon, $Z$ and $W$ boson fields, respectively. These gauge interactions induce pair production of both components of the VLL doublet at a muon collider. The neutral process $\mu^+\mu^-\rightarrow N\bar{N}$ proceeds through $s$-channel $Z$ boson exchange, whereas the charged partner can be produced through $\mu^+\mu^-\rightarrow E^+E^-$ via both $s$-channel photon and $Z$ boson exchanges. Since charged VLL pair production at muon colliders has already been studied in Refs.~\cite{Guo:2023jkz,Morais:2021ead, Ghosh:2023xbj}, we focus on the neutral $N\bar{N}$ channel in this work.

The decays of the VLLs arise from their mixing with the second-generation SM leptons after electroweak symmetry breaking. Assuming that the VLL-SM lepton mixing is small, the relevant interaction Lagrangian responsible for the decays of $E$ and $N$ can be written as
\begin{eqnarray}
\begin{split}
\label{eq:2.7}
\mathcal{L}_{\rm int} =&g_{\bar{\mu}^{\dagger}\bar{N}}^{W^+}[ W_\mu^{+}(\bar{\mu}^{\dagger}\bar{\sigma}^\mu \bar{N}) + W_\mu^{-}(\bar{N}^{\dagger}\bar{\sigma}^\mu \bar{\mu})]
+ g_{\bar{\mu}^{\dagger}\bar{E}}^{Z}Z_\mu (\bar{\mu}^{\dagger}\bar{\sigma}^\mu \bar{E} + \bar{E}^{\dagger}\bar{\sigma}^\mu \bar{\mu})\\
&+ ( y_{E\bar{\mu}}^{h}hE\bar{\mu} + \mathrm{c.c.}),
\end{split}
\end{eqnarray}
where only terms linear in the mixing parameter $\epsilon$ have been retained. Here, $h$ denotes the physical Higgs boson with mass $m_h=125$ GeV. The effective couplings $g_{\bar{\mu}^{\dagger}\bar{N}}^{W^+}$, $g_{\bar{\mu}^{\dagger}\bar{E}}^{Z}$ and $y_{E\bar{\mu}}^{h}$ are given by
\begin{eqnarray}
\begin{split}
\label{eq:2.8}
g_{\bar{\mu}^{\dagger}\bar{N}}^{W^+}=-\epsilon M_W/M_L ,
\end{split}
\end{eqnarray}
\begin{eqnarray}
\begin{split}
\label{eq:2.9}
g_{\bar{\mu}^{\dagger}\bar{E}}^{Z}=-\epsilon M_Z/\sqrt{2}M_L ,
\end{split}
\end{eqnarray}
\begin{eqnarray}
\begin{split}
\label{eq:2.10}
y_{E\bar{\mu}}^{h}=-\epsilon/\sqrt{2} .
\end{split}
\end{eqnarray}

The relevant decay widths of the neutral VLL $N$ are given by
\begin{eqnarray}
\begin{split}
\label{eq:2.11}
\Gamma(N\rightarrow W^+ \mu^-) = \frac{M_N}{32\pi}(1-r_W)^2(2+1/r_W)|g_{\bar{\mu}^{\dagger}\bar{N}}^{W^+}|^2,
\end{split}
\end{eqnarray}
\begin{eqnarray}
\begin{split}
\label{eq:2.12}
\Gamma(N\rightarrow Z \nu) = \Gamma(N\rightarrow h \nu) = 0 ,
\end{split}
\end{eqnarray}
where $r_W = M^2_W/M^2_N$. As can be seen from the decay width formulas in Eqs.~(\ref{eq:2.11})--(\ref{eq:2.12}), the branching ratio of the neutral VLL is
\begin{eqnarray}
\begin{split}
\label{eq:2.13}
BR(N\rightarrow W^+ \mu^-) =  BR(\bar{N}\rightarrow W^- \mu^+) = 1 .
\end{split}
\end{eqnarray}

To motivate our choice of the benchmark value of $\epsilon$, we briefly discuss the constraints on the VLL--SM lepton mixing angle. In the small-mixing limit, the mixing angle is approximately related to $\epsilon$ through
\begin{eqnarray}
\begin{split}
\label{eq:2.14}
\sin\theta_L = \frac{\epsilon v}{M_L} .
\end{split}
\end{eqnarray}
Theoretical considerations, including perturbativity, vacuum stability and constraints from electroweak precision observables, generally favor small VLL--SM lepton mixing angles for TeV-scale VLLs. Current studies indicate that values of $\sin\theta_L\lesssim 0.05$ remain compatible with these constraints~\cite{Cingiloglu:2024vdh,Adhikary:2024esf,Cynolter:2008ea,Lavoura:1992np}.
In this work, we adopt $\epsilon=0.01$ as a benchmark value for the small VLL--SM lepton mixing. For the VLL mass range considered in this work, this choice corresponds to a mixing angle well below the current upper limit. The $N\bar{N}$ pair-production cross section does not depend on $\epsilon$, while $\epsilon$ controls the decay width and lifetime of $N$.
The decay widths are evaluated using the UFO model, and the corresponding widths, proper decay lengths and width-to-mass ratios are summarized in Table~\ref{tab11}.

As shown in Table~\ref{tab11}, $\Gamma_N/M_N$ is much smaller than unity over the entire mass range, validating the narrow-width approximation. The corresponding proper decay lengths $c\tau_N$ are also extremely small, confirming that $N$ decays promptly for $\epsilon=0.01$. Since the $N\mu W$ coupling is proportional to $\epsilon$, the decay width approximately scales as $\Gamma_N \propto \epsilon^2$ and hence $c\tau_N\propto \epsilon^{-2}$. For sufficiently small values of $\epsilon$, $N$ may become long-lived, in which case a dedicated search for displaced-decay signatures would be required. Such long-lived scenarios are beyond the scope of the present work. As mentioned above, the current lower bound on the mass of a doublet vector-like muon is $1270$ GeV~\cite{ATLAS:2024mrr}. Motivated by this limit, we investigate the pair production of neutral doublet VLLs with masses in the range of $1300-2980$ GeV at the muon collider with $\sqrt{s}=6$ TeV.

\begin{table}[H]
\begin{center}
\renewcommand{\arraystretch}{1}
\setlength{\tabcolsep}{20pt}{
\caption{Decay widths, proper decay lengths and width-to-mass ratios of the neutral VLL $N$ for $\epsilon=0.01$.}
\label{tab11}
\resizebox{0.9\textwidth}{!}{
\begin{tabular}{c|ccc}
\hline \hline
    $M_N$ (GeV)  &  $\Gamma_N$ (GeV)  &  $c\tau_N $(mm) & $\Gamma_N/M_N$ \\
 \hline
	1300  &  $1.29\times10^{-3}$  &  $1.53\times10^{-10}$  & $9.93\times10^{-7}$\\
	1700  &  $1.69\times10^{-3}$  &  $1.17\times10^{-10}$  & $9.94\times10^{-7}$ \\
	2100  &  $2.09\times10^{-3}$  &  $9.46\times10^{-11}$  & $9.94\times10^{-7}$ \\
    2500  &  $2.48\times10^{-3}$  &  $7.94\times10^{-11}$  & $9.94\times10^{-7}$ \\ \hline \hline
\end{tabular}}}
\end{center}
\end{table}

%%%%%%%%%%%%%%%%%%%%%%%%%%%%%%%%%%%%%%%%%%%%%%%%%%
%%%%%%%%%%%%%%%%%%%%%%%%%%%%%%%%%%%%%%%%%%%%%%%%%%
\section{Signatures of the neutral doublet VLL at the muon collider}
\begin{figure}[H]
\begin{center}
\centering\includegraphics [scale=0.4] {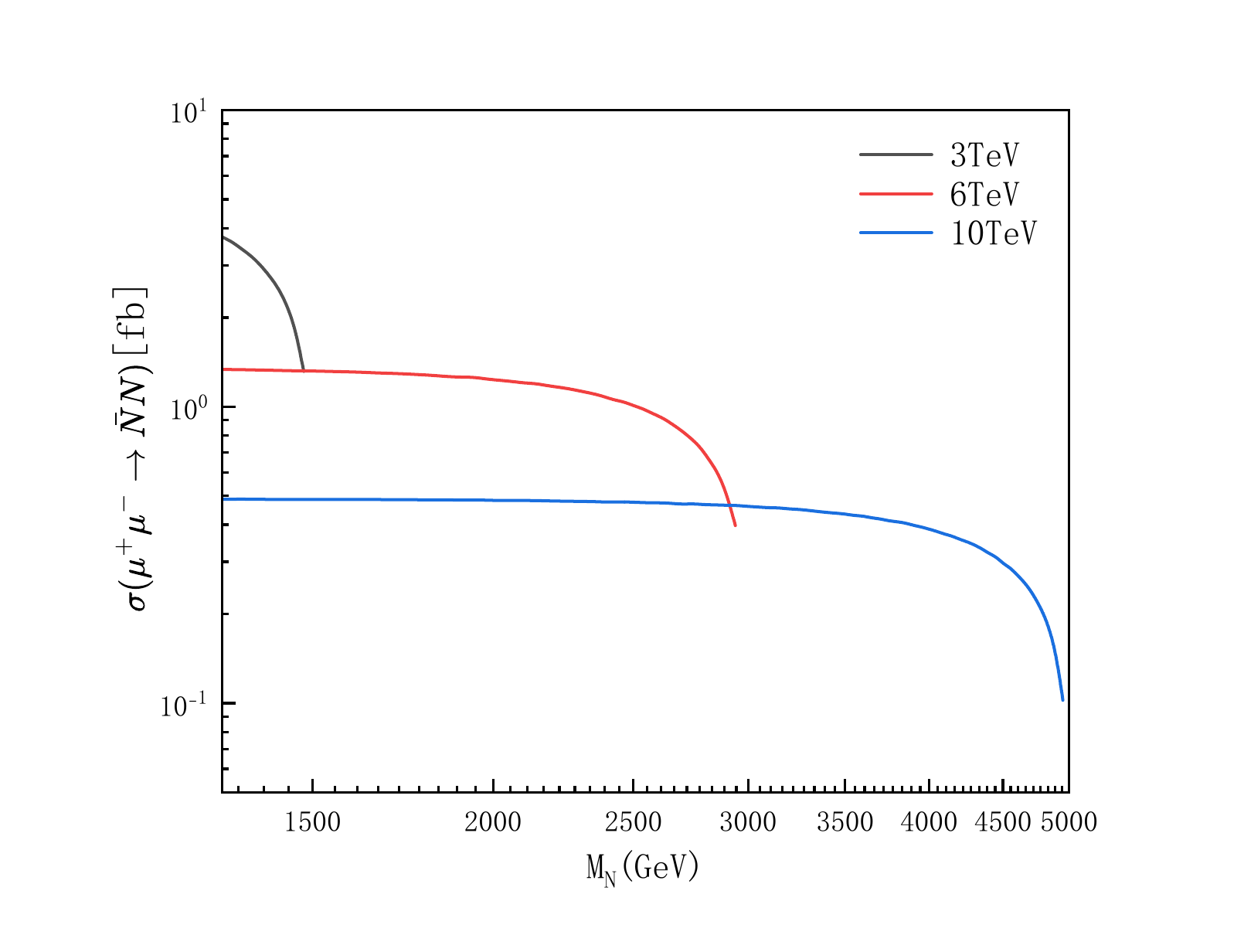}
\caption{The production cross sections for the process $\mu^+\mu^-\rightarrow N\bar{N}$ at center-of-mass energies of $3$, $6$ and $10$ TeV for unpolarized beams.}
\label{fig:1}
\end{center}
\end{figure}

%%%%%%%%%%%%%%%%%%%%%%%%%%%%%%%%%%%%%%%%%%%%%%%%%%
%%%%%%%%%%%%%%%%%%%%%%%%%%%%%%%%%%%%%%%%%%%%%%%%%%
In this section, we investigate the potential of a $6$ TeV muon collider with an integrated luminosity of $4$ ab$^{-1}$ to probe neutral vector-like leptons through pair production. According to Eqs.~(\ref{eq:2.3})--(\ref{eq:2.10}), the process $\mu^+\mu^-\rightarrow N\bar{N}$ proceeds through the exchange of the electroweak gauge boson $Z$.
To place the $6$ TeV benchmark in a broader context, we compare the production cross sections for $\mu^+\mu^-\rightarrow N\bar{N}$ at center-of-mass energies of $3$, $6$ and $10$ TeV, as shown in Fig.~\ref{fig:1}.

The same model parameters and the unpolarized beam configuration are used for the three collider energies, with only the center-of-mass energy varied. For a fixed value of $M_N$, the production cross section generally decreases with increasing center-of-mass energy once the process is sufficiently above threshold. On the other hand, a larger $\sqrt{s}$ extends the kinematically accessible mass range. The $6$ TeV setup therefore provides enough collision energy to access the mass region considered in this work. Although a $10$ TeV collider would extend the accessible mass range further, this production-level comparison alone does not determine the relative search sensitivities at different collider energies, which also depend on the background rates, integrated luminosities, detector performance and event-selection efficiencies. Dedicated searches at $3$ and $10$ TeV would require separate detector-level analyses.

The produced neutral VLL decays via $N\rightarrow W^+ \mu^-$, and both the hadronic $(W\rightarrow jj)$ and leptonic $(W\rightarrow \ell\nu)$ decay channels of the $W$ boson are investigated in the following analysis. Here $\ell = e, \mu$ and $j$ denotes a light jet (u, d, c or s). Therefore, the signals are defined as follows:

\textbullet{ Signal $1$: $\mu^+\mu^-\rightarrow N\bar{N}\rightarrow W^+(\rightarrow jj)W^-(\rightarrow jj)\mu^+\mu^-\rightarrow 4j2\mu$,}

\begin{figure}[H]
\begin{center}
\subfigure[]{\includegraphics [scale=0.4] {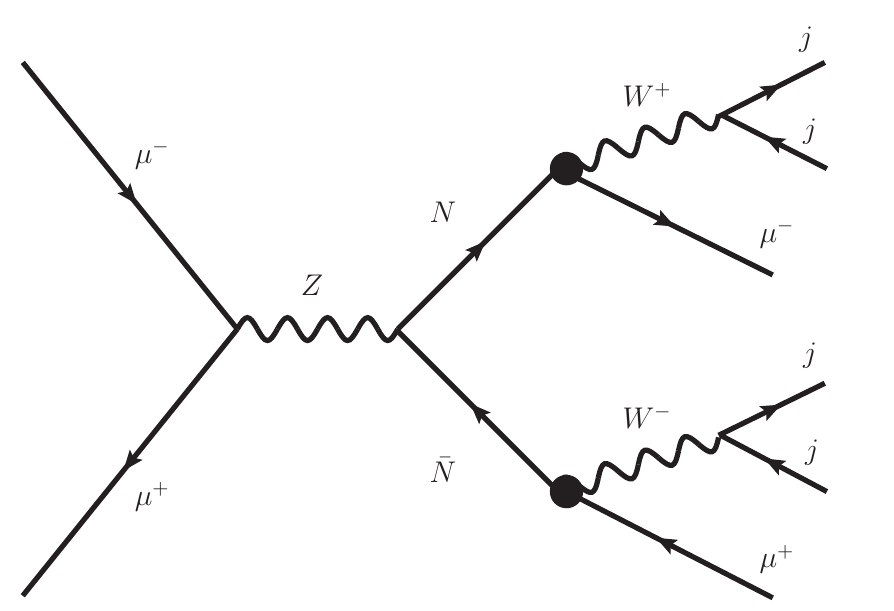}}
\hspace{0.3in}
\subfigure[]{\includegraphics [scale=0.4] {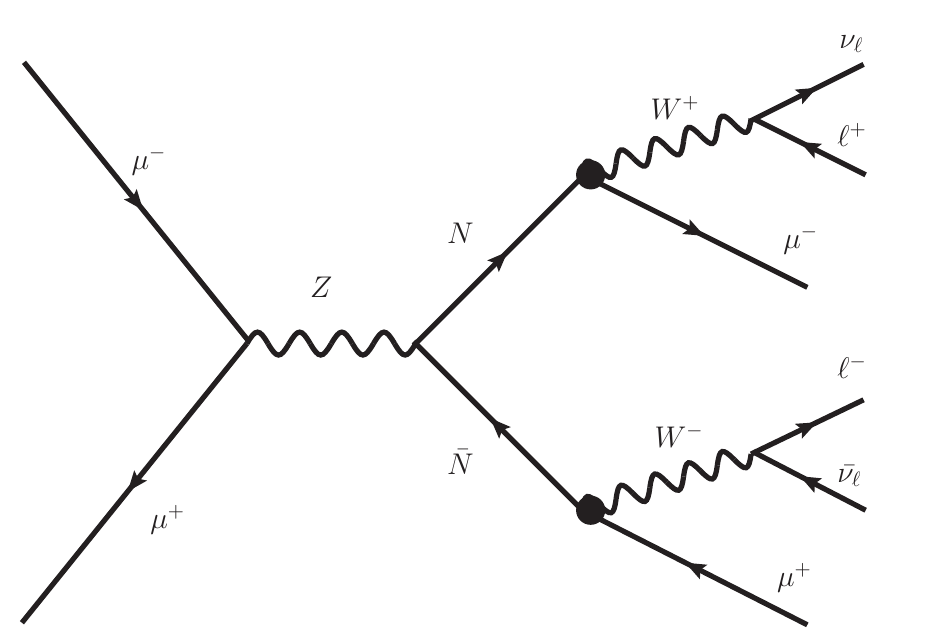}}
\caption{{Feynman diagrams for Signal $1$ (a) and Signal $2$ (b) at the muon collider.}}
\label{fig:2}
\end{center}
\end{figure}

\textbullet{ Signal $2$: $\mu^+\mu^-\rightarrow N\bar{N}\rightarrow W^+(\rightarrow \ell^+\nu_\ell)W^-(\rightarrow \ell^-\bar{\nu_\ell})\mu^+\mu^-\rightarrow 2\ell2\mu E\mkern-10.5 mu/$,}\\
where $E\mkern-10.5 mu/$ represents missing energy. The corresponding Feynman diagrams are shown in Fig.~\ref{fig:2}.

The model is constructed using \verb"FeynRules"~\cite{Alloul:2013bka}, which is then used to generate the corresponding model files in the UFO format~\cite{Degrande:2011ua}. Monte Carlo simulations are then performed to study the pair production of $N$ at the muon collider. Both the signal and the relevant SM background events are generated with \verb"MadGraph5_aMC@NLO"~\cite{Alwall:2014hca} version $3.6.6$. We generate $10^5$ Monte Carlo events for each signal benchmark point and $3\times10^5$ events for each SM background sample.
In the following subsections, the signal and background processes are analyzed with the following loose parton-level generation cuts:

For Signal $1$:
\[
\begin{aligned}
& p_T^\ell > 10 \:\text{GeV}, \quad |\eta_\ell| < 2.5, \\
& p_T^j > 20 \:\text{GeV}, \quad |\eta_j| < 5, \\
& \Delta R_{\ell\ell} > 0.4, \quad \Delta R_{jj} > 0.4, \quad \Delta R_{\ell j} > 0.4.
\end{aligned}
\]

For Signal $2$:
\[
p_T^\ell > 10 \:\text{GeV},\quad  |\eta_\ell| < 2.5,\quad  \Delta R_{\ell\ell} > 0.4.
\]
Here $p_T^\ell$ and $p_T^j$ denote the transverse momenta of leptons and jets. $|\eta_\ell|$ and $|\eta_j|$ represent the absolute values of the pseudorapidities of leptons and jets. $\Delta R_{\ell\ell}$, $\Delta R_{jj}$ and $\Delta R_{\ell j}$ refer to the angular separations between two leptons, two jets, and a lepton and a jet, respectively. Here, $\Delta R$ is defined as $\sqrt{(\Delta \phi)^2+(\Delta \eta)^2}$.

\begin{figure}[H]
\begin{center}
\subfigure[]{\includegraphics [scale=0.27] {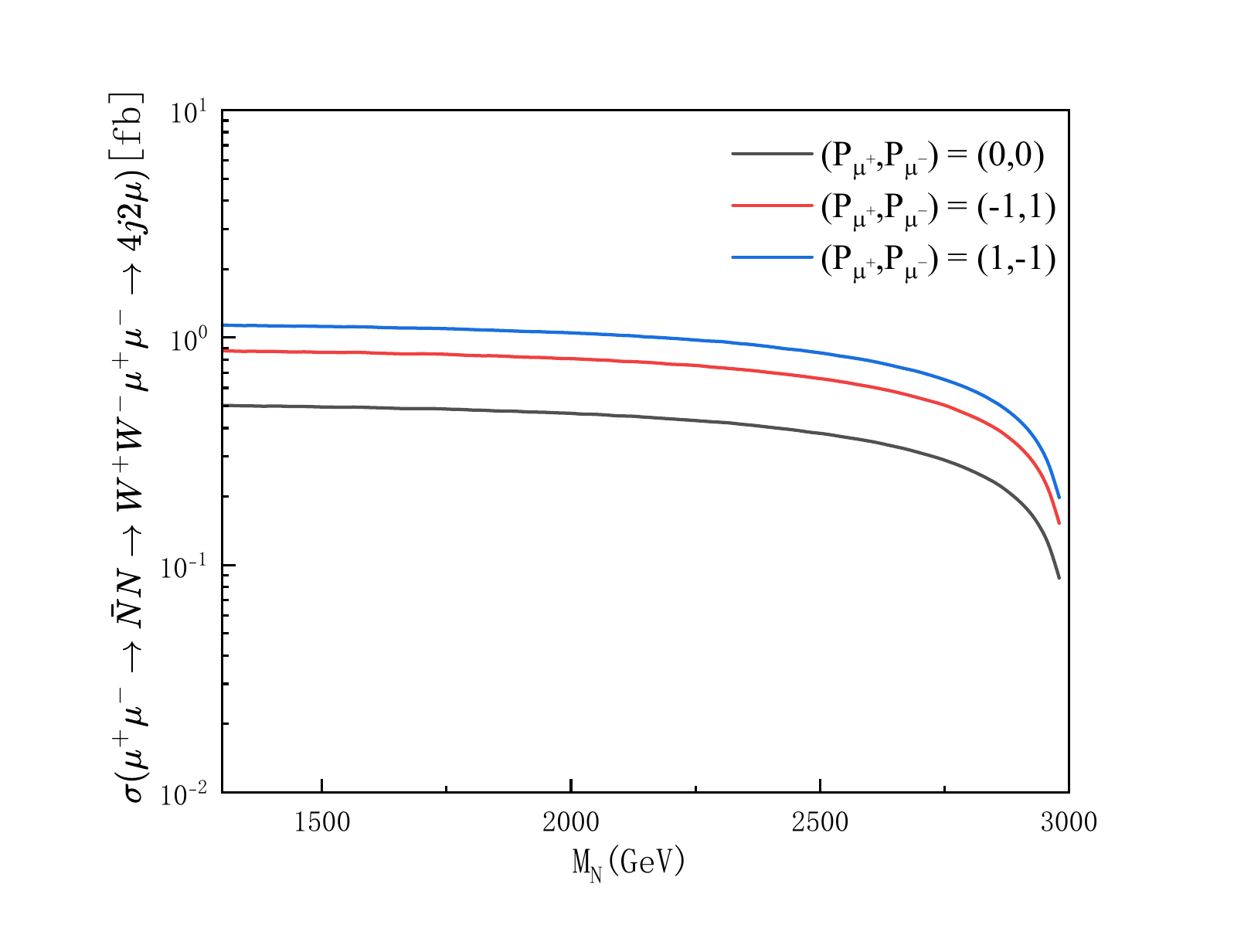}}
\subfigure[]{\includegraphics [scale=0.27] {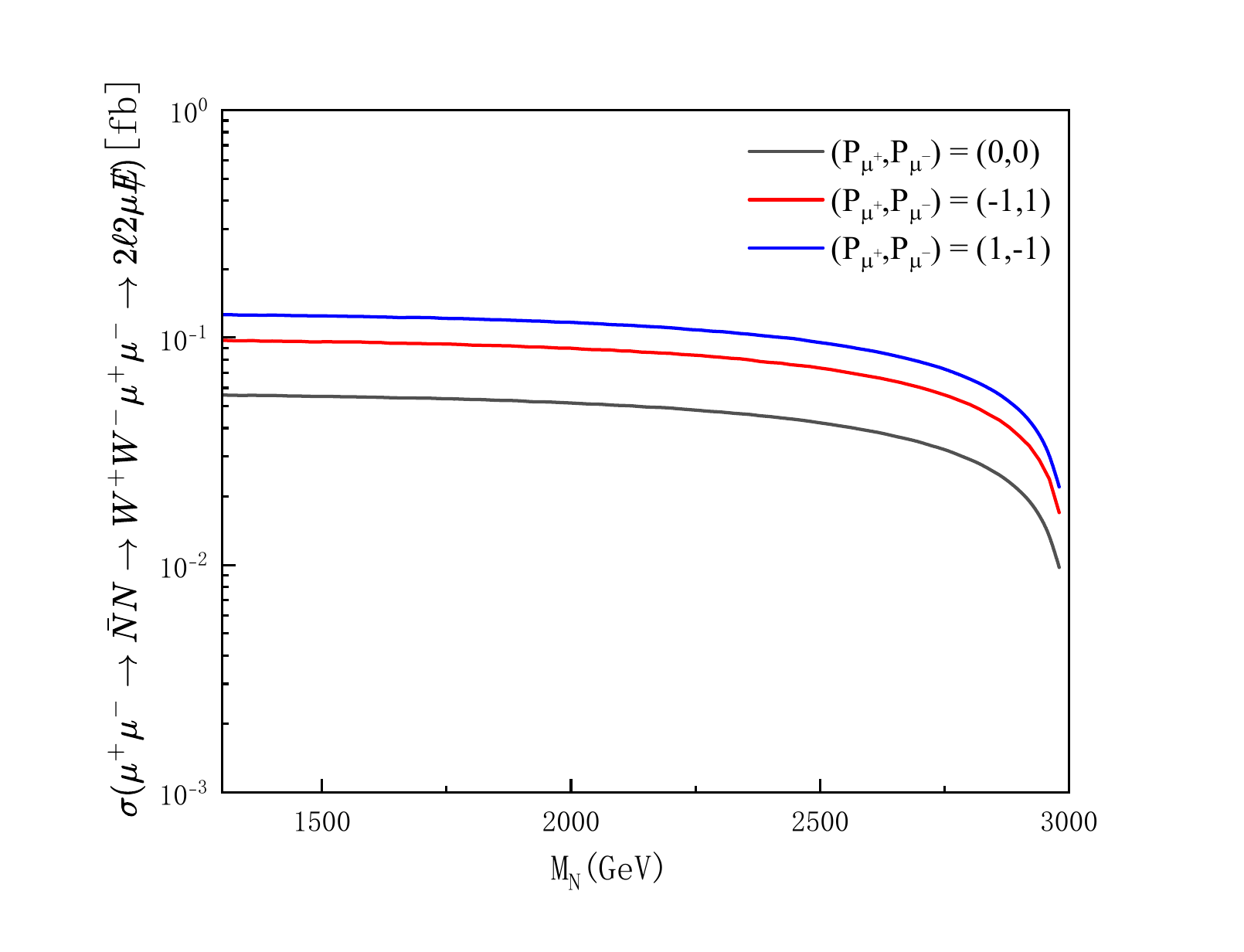}}
\caption{{The cross sections for Signal $1$ (a) and Signal $2$ (b) at the $6$ TeV muon collider.}}
\label{fig:3}
\end{center}
\end{figure}

For a muon collider, beam polarizations $P_{\mu^+}$ and $P_{\mu^-}$ provide an additional handle to enhance the sensitivity to new physics signals. We adopt the convention $P=(N_R-N_L)/(N_R+N_L)$, where $N_R$ and $N_L$ denote the numbers of right-handed and left-handed beam particles, respectively. Thus, $P =+1(-1)$ corresponds to a fully right-handed (left-handed) beam, while $P=0$ denotes an unpolarized beam.
In the notation $(P_{\mu^+},P_{\mu^-})$, the first and second entries refer to the $\mu^+$ and $\mu^-$ beams, respectively.
The production cross sections of Signal $1$ and $2$ for different beam polarization configurations are shown in Fig.~\ref{fig:3}. The configurations $(P_{\mu^+},P_{\mu^-})=(1,1)$ and $(-1,-1)$ lead to highly suppressed cross sections and are therefore neglected in the following discussions. For the remaining two polarization configurations, the $(P_{\mu^+},P_{\mu^-})=(1,-1)$ case generally gives a larger signal cross section than the $(P_{\mu^+},P_{\mu^-})=(-1,1)$ case.

This polarization dependence of the signal cross sections can be understood from the helicity and chiral structure of the $Z$-mediated production process. For highly relativistic initial-state muons, the $s$-channel process is dominated by opposite-helicity initial states, whereas same-helicity initial states are strongly suppressed. Under our convention, the $(P_{\mu^+},P_{\mu^-})=(1,-1)$ and $(-1,1)$ configurations probe the left- and right-handed $Z$-muon couplings $g_L^\mu$ and $g_R^\mu$, respectively.
Since $\mid g_L^\mu\mid > \mid g_R^\mu\mid$, the $(P_{\mu^+},P_{\mu^-})=(1,-1)$ configuration gives a somewhat larger signal cross section than the $(P_{\mu^+},P_{\mu^-})=(-1,1)$ configuration. Therefore, if only the signal rate is considered, $(P_{\mu^+},P_{\mu^-})=(1,-1)$ would be the preferred polarization configuration.

However, the preferred beam polarization should be determined by considering both the signal and the SM background rates. We therefore evaluate the total background cross sections for the same polarization configurations, with the individual background processes discussed separately in the following channel analyses. The results are summarized in Table~\ref{tab10}. The stronger polarization dependence of the backgrounds mainly comes from the chiral structure of the electroweak interactions, especially charged-current processes involving $W$ bosons.
Since the $W$ boson couples only to left-handed lepton fields, the charged-current interaction selects left-handed $\mu^-$ and right-handed $\mu^+$ beams in the high-energy limit. Therefore, the $(P_{\mu^+},P_{\mu^-})=(1,-1)$ configuration tends to enhance these backgrounds, whereas the $(P_{\mu^+},P_{\mu^-})=(-1,1)$ configuration suppresses them more. Thus, although the $(P_{\mu^+},P_{\mu^-})=(1,-1)$ configuration gives a larger signal rate, it also leads to much larger background rates. In contrast, the $(P_{\mu^+},P_{\mu^-})=(-1,1)$ configuration retains a sizable signal rate while providing stronger background suppression. It therefore gives a better signal-to-background ratio and statistical significance, and is adopted in the following polarized analysis.
The fully polarized case is used only as an idealized benchmark to illustrate the maximal polarization effect, since achieving full beam polarization at a high-energy muon collider may be experimentally challenging. We therefore also present the unpolarized results as a reference.

\begin{table}[H]
\begin{center}
\renewcommand{\arraystretch}{1}
\setlength{\tabcolsep}{24pt}{
\caption{Total background cross sections for the $4j2\mu$ (Signal $1$) and $2\ell2\mu E\mkern-10.5 mu/$ (Signal $2$) signals under different beam polarization configurations after applying the basic cuts.}
\label{tab10}
\resizebox{0.9\textwidth}{!}{
\begin{tabular}{c|cc}
\hline \hline
    $(P_{\mu^+},P_{\mu^-})$  &  Background for Signal $1$ [fb] &  Background for Signal $2$ [fb]  \\
 \hline
	(0,0)   &  $7.46$   &  $1.50$   \\
	(1,-1)  &  $26.64$  &  $5.10$   \\
	(-1,1)  &  $2.48$   &  $0.57$   \\ \hline \hline
\end{tabular}}}
\end{center}
\end{table}

The signal and SM background events are then passed to \verb"PYTHIA8"~\cite{Sjostrand:2014zea} version $8.317$ for parton showering and hadronization. Detector effects are simulated using \verb"DELPHES"~\cite{deFavereau:2013fsa} version $3.5.0$ with the ``MuonCollider" detector card~\cite{MuonColliderDelphesCard} distributed with this version. This card provides a parametrization of the target performance of a muon collider detector and does not represent a full detector simulation. In this setup, both the electromagnetic and hadronic calorimeters cover the range $|\eta|< 2.5$ and reconstructed jets are restricted to the same acceptance range.
Jet reconstruction is performed using the Valencia (VLC) algorithm~\cite{Boronat:2014hva,Boronat:2016tgd}, with an exclusive $R = 0.5$ and $N = 2$ configuration, corresponding to the ``VLCjetR05N2" collection in the ``MuonCollider" card. For the $4j2\mu$ signal, defined according to its parton-level decay topology, the four quarks from the two hadronically decaying $W$ bosons are clustered into two reconstructed VLC jets, while the muons are reconstructed separately.
Beam-induced backgrounds from muon decays are not explicitly simulated in the present Delphes-based fast simulation. Therefore, the quoted sensitivities should be interpreted as fast-simulation projections based on the target detector performance assumed in the ``MuonCollider" card, rather than as full-simulation projections. Finally, kinematic and cut-based analyses are performed
using \verb"MadAnalysis5"~\cite{Conte:2012fm,Conte:2014zja,Conte:2018vmg} version $1.9.60$. The details of the signal and background processes are discussed in the following subsections.

\subsection{The $4j2\mu$ signal}
For the $4j2\mu$ signal, the signal cross section decreases from $0.87$ fb to $0.15$ fb as $M_N$ increases from $1300$ to $2980$ GeV for the polarized configuration $(P_{\mu^+},P_{\mu^-})=(-1,1)$. For comparison, the corresponding unpolarized cross section decreases from $0.50$ fb to $8.75\times 10^{-2}$ fb over the same mass range. The main SM background processes considered are:
(I) $\mu^+\mu^-\rightarrow \mu^+\mu^-jjjj$, including the complete tree-level electroweak contribution to this final state, with all relevant diagrams and interference effects;
(II) $\mu^+\mu^-\rightarrow t\bar{t}$, with both top quarks decaying hadronically;
(III) $\mu^+\mu^-\rightarrow W^+W^-jj$, with both $W$ bosons decaying hadronically.
Although the latter two processes do not contain two prompt muons at the matrix-element level, muons can arise from heavy-flavour hadron decays after parton showering and hadronization and may subsequently be reconstructed in the Delphes simulation. A small fraction of these events can therefore satisfy the requirement of at least two reconstructed muons.
These contributions are included in the analysis and subjected to the same selection cuts as the other backgrounds. All reconstructed jets are treated inclusively, without distinguishing between $b$-jets and light-flavour jets. No separate fake-object background model is included beyond the object reconstruction implemented in Delphes. Given the very small residual $t\bar{t}$ contribution after the final selections, we do not introduce a $b$-tag veto in this analysis.

\begin{figure}[H]
\begin{center}
\subfigure[]{\includegraphics [scale=0.33] {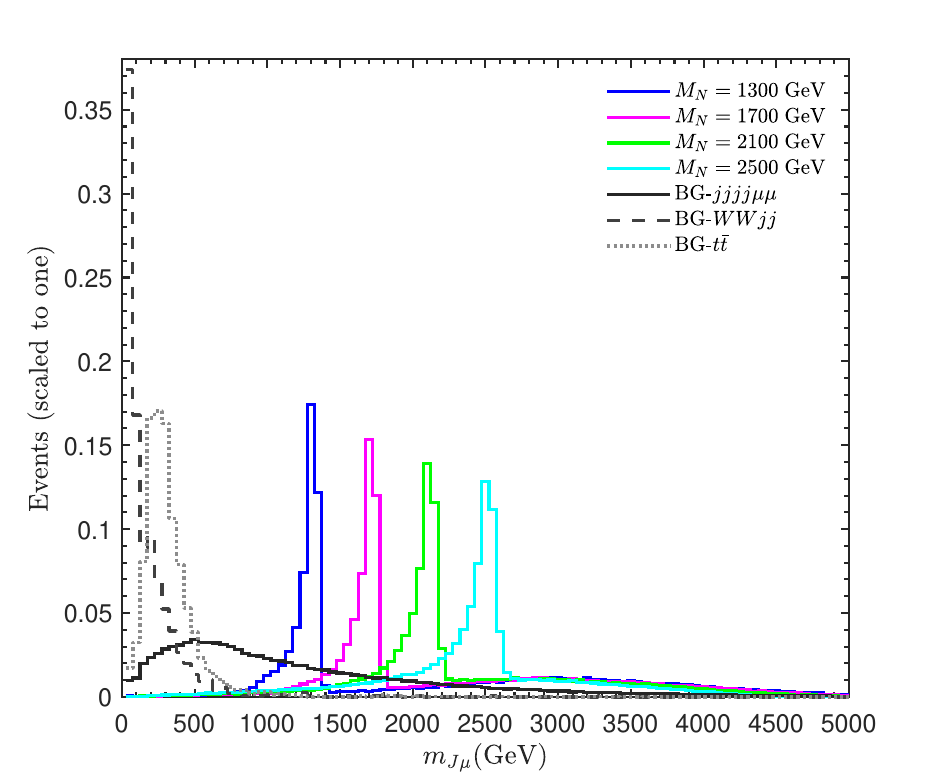}}
\subfigure[]{\includegraphics [scale=0.33] {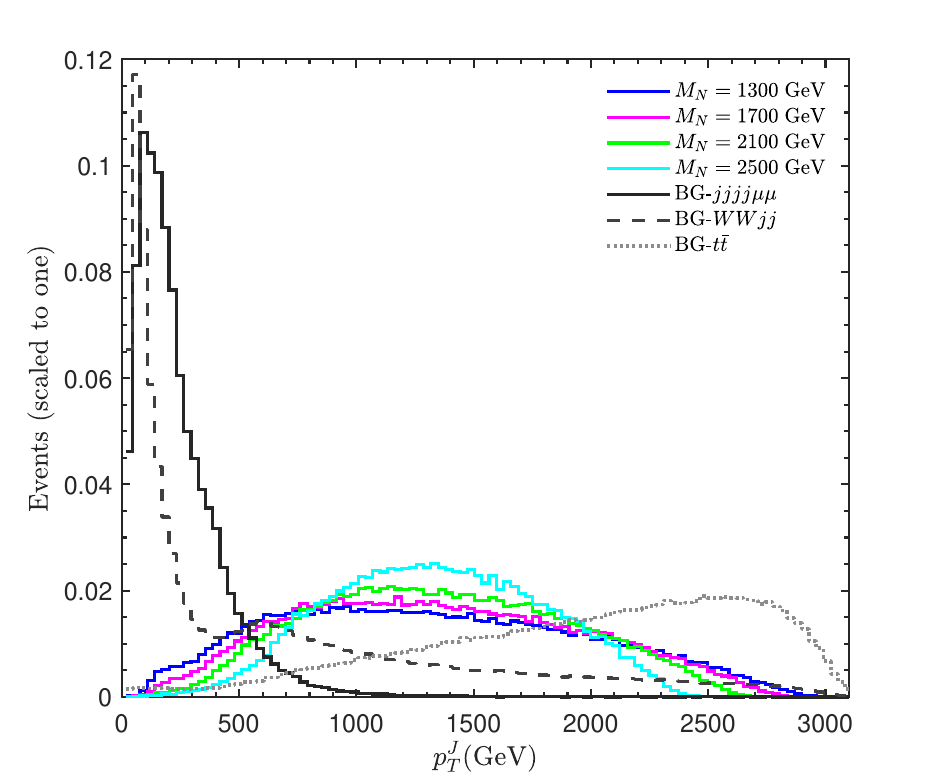}}
\subfigure[]{\includegraphics [scale=0.33] {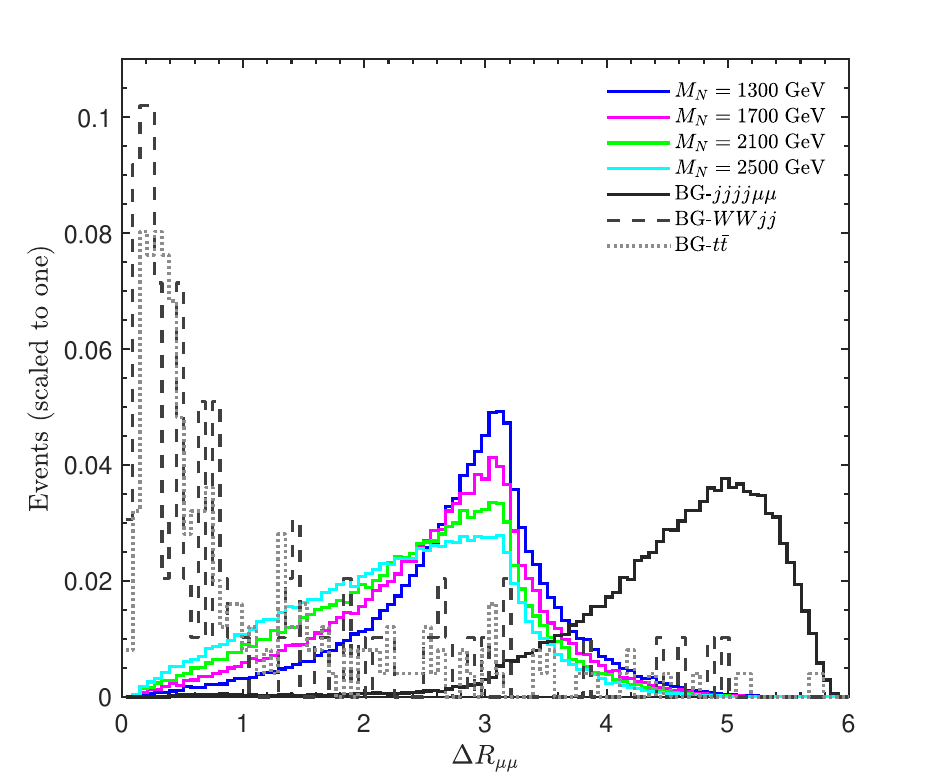}}
\vspace{-0.3cm}
\caption{The normalized distributions of the observables $m_{J\mu}$ (a), $p_T^{J}$ (b) and $\Delta R_{\mu\mu}$ (c) for the signals of selected neutral VLL-mass benchmark points and the SM backgrounds at the $6$ TeV muon collider with $\mathcal{L}=$ $4$ ab$^{-1}$ for the unpolarized beam configuration $(P_{\mu^+},P_{\mu^-})=(0,0)$.}
\label{fig:4}
\end{center}
\end{figure}

\begin{figure}[H]
\begin{center}
\subfigure[]{\includegraphics [scale=0.33] {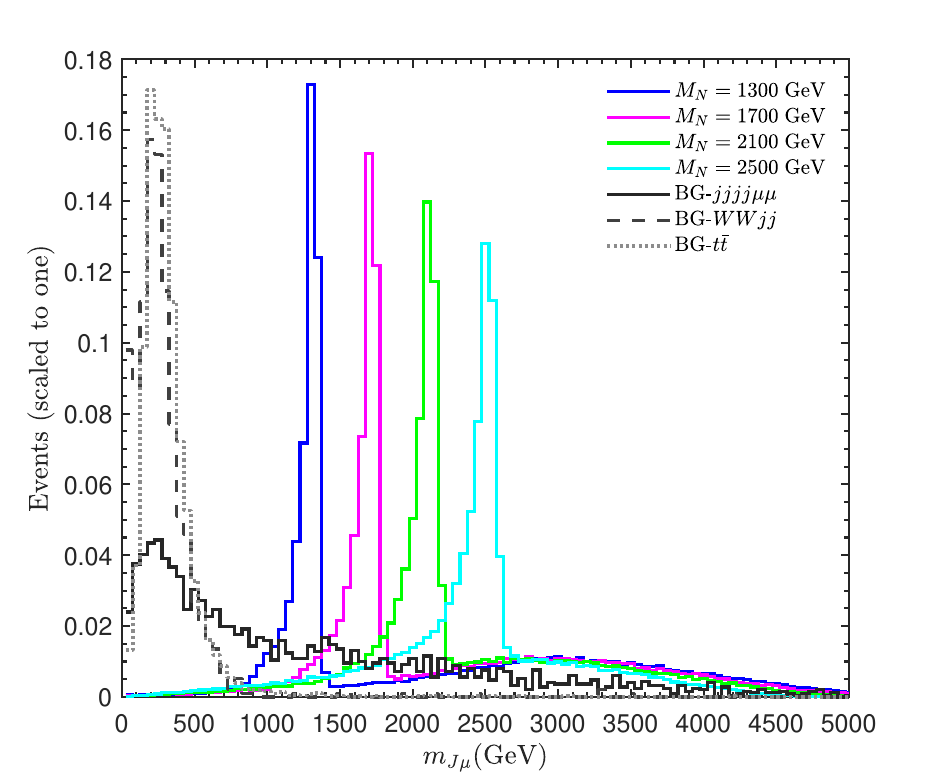}}
\subfigure[]{\includegraphics [scale=0.33] {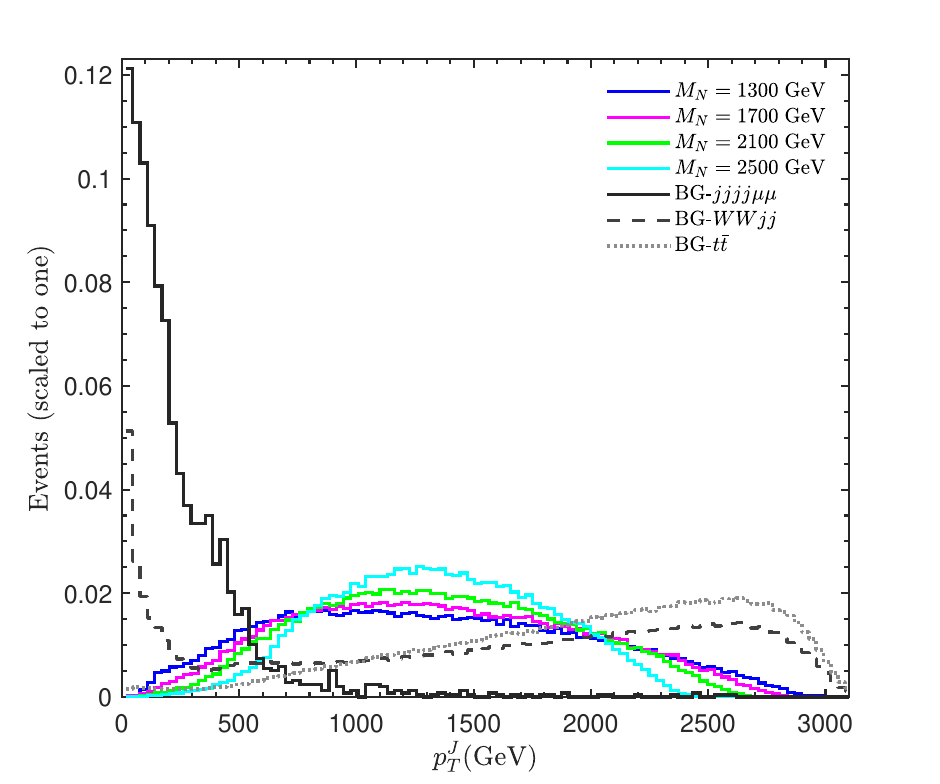}}
\subfigure[]{\includegraphics [scale=0.33] {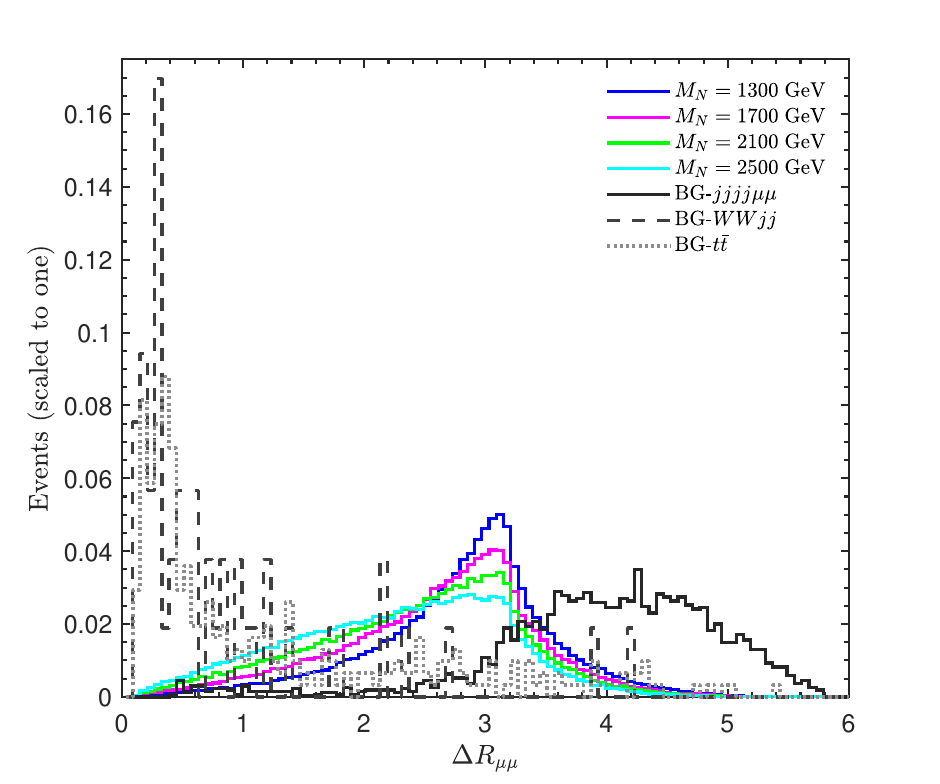}}
\vspace{-0.3cm}
\caption{Same as Fig.~\ref{fig:4} but for the polarized beam configuration $(P_{\mu^+},P_{\mu^-})=(-1,1)$.}
\label{fig:5}
\end{center}
\end{figure}

To separate the $4j2\mu$ signal from the SM backgrounds, we first require at least two reconstructed muons, $N_\mu\geq2$, as expected from the signal final state. This requirement strongly suppresses backgrounds such as $t\bar t$ and $WWjj$, which contain a much smaller fraction of events with two selected muons.
We then consider the invariant mass, $m_{J\mu}$, the transverse momentum, $p_T^J$ and the angular separation between the two muons, $\Delta R_{\mu\mu}$, to further separate the signal from the remaining backgrounds. Here, $J$ denotes the highest-$p_T$ reconstructed VLC jet, and $m_{J\mu}=m(J,\mu^-)$ is constructed from $J$ and the highest-$p_T$ negatively charged reconstructed muon. After applying the basic selection cuts, the normalized distributions of these variables at the $6$ TeV muon collider with $\mathcal{L}=4$ ab$^{-1}$ are shown in Figs.~\ref{fig:4} and~\ref{fig:5} for the unpolarized and polarized beam configurations, respectively. The colored solid lines correspond to the signal benchmark points $M_N=1300$, $1700$, $2100$, and $2500$ GeV, while the other curves represent the different SM background processes. Although the relative contributions and detailed shapes of the SM backgrounds change with the beam polarization, the main kinematic features that distinguish the signal from the backgrounds remain similar in the two configurations.

Since the neutral VLL decays through $N\to W\mu$, $m_{J\mu}$ is used as a reconstructed $W\mu$ invariant-mass observable. As shown in Figs.~\ref{fig:4}(a) and~\ref{fig:5}(a), the signal distributions peak near the corresponding values of $M_N$, whereas the SM backgrounds have no such heavy resonance and are mainly distributed at lower masses or over a broader range.
The transverse momentum $p_T^J$ provides complementary discrimination. In the signal, the reconstructed hadronic $W$ bosons originate from the two-body decay of a TeV-scale VLL, and therefore typically carry a hard momentum scale set by $M_N$. Consequently, the signal tends to populate relatively large values of $p_T^J$. In contrast, the SM backgrounds do not contain a heavy resonance decaying into $W\mu$, and important background contributions are more concentrated toward lower $p_T^J$, as shown in Figs.~\ref{fig:4}(b) and~\ref{fig:5}(b). We also consider the angular separation $\Delta R_{\mu\mu}$ between the two muons. For the signal, the two muons originate from the decays of two separate heavy VLLs, while the SM backgrounds arise from different production and decay topologies, leading to different angular correlations. The signal is mainly concentrated at intermediate values of $\Delta R_{\mu\mu}$, whereas part of the background extends to larger values, as shown in Figs.~\ref{fig:4}(c) and~\ref{fig:5}(c).

\begin{table}[H]
\begin{center}
\renewcommand{\arraystretch}{1}
\setlength{\tabcolsep}{28pt}{
\caption{The improved cuts on the signal and SM backgrounds for the $4j2\mu$ signal.}
\label{tab1}
\resizebox{0.9\textwidth}{!}{
\begin{tabular}[c]{c|c|c}
\hline\hline
 & $(P_{\mu^+},P_{\mu^-})=(0,0)$ & $(P_{\mu^+},P_{\mu^-})=(-1,1)$ \\
\hline
\multirow{3}{*}{Basic cuts}
& $p_T^\ell > 10~\mathrm{GeV},\ |\eta_\ell| < 2.5$ & $p_T^\ell > 10~\mathrm{GeV},\ |\eta_\ell| < 2.5$\\
& $p_T^j > 20~\mathrm{GeV},\ |\eta_j| < 5$ & $p_T^j > 20~\mathrm{GeV},\ |\eta_j| < 5$\\
& $\Delta R_{\ell\ell},\ \Delta R_{jj},\ \Delta R_{\ell j} > 0.4$ & $\Delta R_{\ell\ell},\ \Delta R_{jj},\ \Delta R_{\ell j} > 0.4$\\
\hline
Cut 1
& $ N_\mu \geq 2 $
& $N_\mu \geq 2 $ \\
Cut 2
& $m_{J\mu} > 1200~\mathrm{GeV}$
& $m_{J\mu} > 1000~\mathrm{GeV}$ \\
Cut 3
& $p_T^{J} > 500~\mathrm{GeV}$
& -- \\
Cut 4
& $\Delta R_{\mu\mu} < 3.8$
& -- \\
\hline\hline
\end{tabular}}}
\end{center}
\end{table}

Since the relative background contributions differ between the unpolarized and polarized beam configurations, the selections are optimized separately. In the unpolarized case, further kinematic cuts are needed to suppress the remaining background, while in the polarized case the background is already sufficiently reduced after the $N_\mu$ and $m_{J\mu}$ requirements. Additional cuts on $p_T^J$ and $\Delta R_{\mu\mu}$ do not improve the statistical significance and are therefore not applied. The final selection cuts are summarized in Table~\ref{tab1}.

\begin{table}[H]
\scriptsize
\renewcommand{\arraystretch}{1.3}
\setlength{\tabcolsep}{16pt}
\centering
\caption{The cross sections and corresponding expected numbers of events for the signal and SM backgrounds in the unpolarized beam configuration at the $6$ TeV muon collider with $\mathcal{L}=4$ ab$^{-1}$. }
\label{tab2}
\resizebox{0.95\textwidth}{!}{
\begin{tabular}{c|ccccc}
\hline\hline
\multirow{2}{*}{Process}
& \multicolumn{5}{c}{cross sections for signal and backgrounds [fb] (expected events)} \\
\cline{2-6}
& Basic cuts
& Cut 1
& Cut 2
& Cut 3
& Cut 4 \\
\hline

$M_N=1300~\mathrm{GeV}$
& \makecell{$5.02\times10^{-1}$\\$(2009)$}
& \makecell{$4.44\times10^{-1}$\\$(1777)$}
& \makecell{$4.37\times10^{-1}$\\$(1746)$}
& \makecell{$3.96\times10^{-1}$\\$(1583)$}
& \makecell{$3.73\times10^{-1}$\\$(1494)$} \\

\hline

$M_N=1700~\mathrm{GeV}$
& \makecell{$4.87\times10^{-1}$\\$(1947)$}
& \makecell{$4.29\times10^{-1}$\\$(1717)$}
& \makecell{$4.27\times10^{-1}$\\$(1710)$}
& \makecell{$3.96\times10^{-1}$\\$(1584)$}
& \makecell{$3.78\times10^{-1}$\\$(1514)$} \\

\hline

$M_N=2100~\mathrm{GeV}$
& \makecell{$4.53\times10^{-1}$\\$(1812)$}
& \makecell{$4.01\times10^{-1}$\\$(1602)$}
& \makecell{$4.00\times10^{-1}$\\$(1600)$}
& \makecell{$3.82\times10^{-1}$\\$(1529)$}
& \makecell{$3.70\times10^{-1}$\\$(1479)$} \\

\hline

$M_N=2500~\mathrm{GeV}$
& \makecell{$3.79\times10^{-1}$\\$(1516)$}
& \makecell{$3.34\times10^{-1}$\\$(1336)$}
& \makecell{$3.34\times10^{-1}$\\$(1334)$}
& \makecell{$3.26\times10^{-1}$\\$(1304)$}
& \makecell{$3.18\times10^{-1}$\\$(1272)$} \\

\hline

$BG-jjjj\mu\mu$
& \makecell{$3.03$\\$(12105)$}
& \makecell{$2.84$\\$(11353)$}
& \makecell{$2.01$\\$(8052)$}
& \makecell{$2.10\times10^{-1}$\\$(840)$}
& \makecell{$6.05\times10^{-2}$\\$(242)$} \\

\hline

$BG-WWjj$
& \makecell{$2.48$\\$(9921)$}
& \makecell{$9.35\times10^{-4}$\\$(3.74)$}
& \makecell{$8.28\times10^{-6}$\\$(3.31\times10^{-2})$}
& \makecell{$8.28\times10^{-6}$\\$(3.31\times10^{-2})$}
& \makecell{$8.28\times10^{-6}$\\$(3.31\times10^{-2})$} \\

\hline

$BG-t\bar{t}$
& \makecell{$1.95$\\$(7815)$}
& \makecell{$2.15\times10^{-3}$\\$(8.60)$}
& \makecell{$1.96\times10^{-5}$\\$(7.82\times10^{-2})$}
& \makecell{$1.96\times10^{-5}$\\$(7.82\times10^{-2})$}
& \makecell{$1.30\times10^{-5}$\\$(5.21\times10^{-2})$} \\

\hline\hline
\end{tabular}
}
\end{table}

\begin{table}[H]
\scriptsize
\renewcommand{\arraystretch}{1.3}
\setlength{\tabcolsep}{28pt}
\centering
\caption{Same as Table~\ref{tab2} but for the polarized beam configuration $(P_{\mu^+},P_{\mu^-})=(-1,1)$.}
\label{tab3}
\resizebox{0.85\textwidth}{!}{
\begin{tabular}{c|ccc}
\hline\hline
\multirow{2}{*}{Process}
& \multicolumn{3}{c}{cross sections for signal and backgrounds [fb] (expected events)} \\
\cline{2-4}
& Basic cuts
& Cut 1
& Cut 2 \\
\hline

$M_N=1300~\mathrm{GeV}$
& \makecell{$8.73\times10^{-1}$\\$(3493)$}
& \makecell{$7.71\times10^{-1}$\\$(3084)$}
& \makecell{$7.66\times10^{-1}$\\$(3066)$} \\

\hline

$M_N=1700~\mathrm{GeV}$
& \makecell{$8.47\times10^{-1}$\\$(3388)$}
& \makecell{$7.47\times10^{-1}$\\$(2987)$}
& \makecell{$7.46\times10^{-1}$\\$(2982)$} \\

\hline

$M_N=2100~\mathrm{GeV}$
& \makecell{$7.88\times10^{-1}$\\$(3152)$}
& \makecell{$6.96\times10^{-1}$\\$(2783)$}
& \makecell{$6.95\times10^{-1}$\\$(2779)$} \\

\hline

$M_N=2500~\mathrm{GeV}$
& \makecell{$6.60\times10^{-1}$\\$(2640)$}
& \makecell{$5.82\times10^{-1}$\\$(2329)$}
& \makecell{$5.82\times10^{-1}$\\$(2326)$} \\

\hline

$BG-jjjj\mu\mu$
& \makecell{$1.45\times10^{-2}$\\$(58)$}
& \makecell{$1.38\times10^{-2}$\\$(55)$}
& \makecell{$9.42\times10^{-3}$\\$(38)$} \\

\hline

$BG-WWjj$
& \makecell{$1.33\times10^{-2}$\\$(53)$}
& \makecell{$5.48\times10^{-6}$\\$(2.19\times10^{-2})$}
& \makecell{$1.71\times10^{-7}$\\$(6.84\times10^{-4})$} \\

\hline

$BG-t\bar{t}$
& \makecell{$2.45$\\$(9804)$}
& \makecell{$3.38\times10^{-3}$\\$(14)$}
& \makecell{$4.90\times10^{-5}$\\$(1.96\times10^{-1})$} \\

\hline\hline
\end{tabular}
}
\end{table}

Tables~\ref{tab2} and~\ref{tab3} summarize the signal and SM background cross sections after applying the basic and improved selection cuts for the unpolarized and polarized beam configurations, respectively, together with the corresponding expected numbers of events for $\mathcal{L} = 4~\mathrm{ab}^{-1}$.
The statistics-only significance $Z_{\rm stat}$ is evaluated as $Z_{\rm stat}= \sqrt{2\left[(S+B)\ln\left(1+\frac{S}{B}\right)-S\right]}$~\cite{Cowan:2010js} ,
where $S$ and $B$ are the expected numbers of signal and background events after the final selection cuts, respectively. This expression assumes that the residual background is known exactly. To estimate the effect of background systematic uncertainties, we introduce a fractional background-normalization uncertainty $\delta_B$. The background normalization is treated as a nuisance parameter with a Gaussian constraint whose standard deviation is set to $\delta_B B$.
The significance including the background-normalization uncertainty is evaluated within the profile-likelihood Asimov framework using Eqs.~$(91)$ and $(92)$ of Ref.~\cite{Basso:2021dwp}.
Since a full study of experimental systematic uncertainties at a future muon collider is beyond the scope of this work, we only consider representative values of $\delta_B=5\%,10\%$ and $20\%$. The resulting significances are summarized in Table~\ref{tab4} for several benchmark values of $M_N$.

\begin{table}[H]
\scriptsize
\renewcommand{\arraystretch}{1.3}
\setlength{\tabcolsep}{8pt}
\centering
\caption{Expected significances for several benchmark values of $M_N$ at the $6$ TeV muon collider with $\mathcal{L}=4~$ ab$^{-1}$, for the unpolarized and polarized beam configurations. Here, $Z_{\rm stat}$ denotes the statistics-only significance, while $Z_{\delta_B=5\%}$, $Z_{\delta_B=10\%}$ and $Z_{\delta_B=20\%}$ include the corresponding fractional background-normalization uncertainties.}
\label{tab4}
\resizebox{0.95\textwidth}{!}{
\begin{tabular}{c|cccc|cccc}
\hline\hline
\multirow{2}{*}{$M_N$ [GeV]}
& \multicolumn{4}{c|}{$(P_{\mu^+},P_{\mu^-})=(0,0)$}
& \multicolumn{4}{c}{$(P_{\mu^+},P_{\mu^-})=(-1,1)$} \\
\cline{2-9}
& $Z_{\rm stat}$
& $Z_{\delta_B=5\%}$
& $Z_{\delta_B=10\%}$
& $Z_{\delta_B=20\%}$
& $Z_{\rm stat}$
& $Z_{\delta_B=5\%}$
& $Z_{\delta_B=10\%}$
& $Z_{\delta_B=20\%}$ \\
\hline

1300
& $62.04$ & $47.57$ & $36.55$ & $24.48$
& $145.65$ & $126.86$ & $113.88$ & $97.91$ \\

\hline

1700
& $62.67$ & $48.05$ & $36.94$ & $24.77$
& $143.12$ & $124.74$ & $111.95$ & $96.19$ \\

\hline

2100
& $61.57$ & $47.22$ & $36.26$ & $24.27$
& $136.80$ & $119.43$ & $107.12$ & $91.89$ \\

\hline

2500
& $54.83$ & $42.12$ & $32.14$ & $21.27$
& $122.04$ & $106.99$ & $95.81$ & $81.84$ \\

\hline\hline
\end{tabular}
}
\end{table}

\begin{figure}[H]
\begin{center}
\centering\includegraphics [scale=0.36] {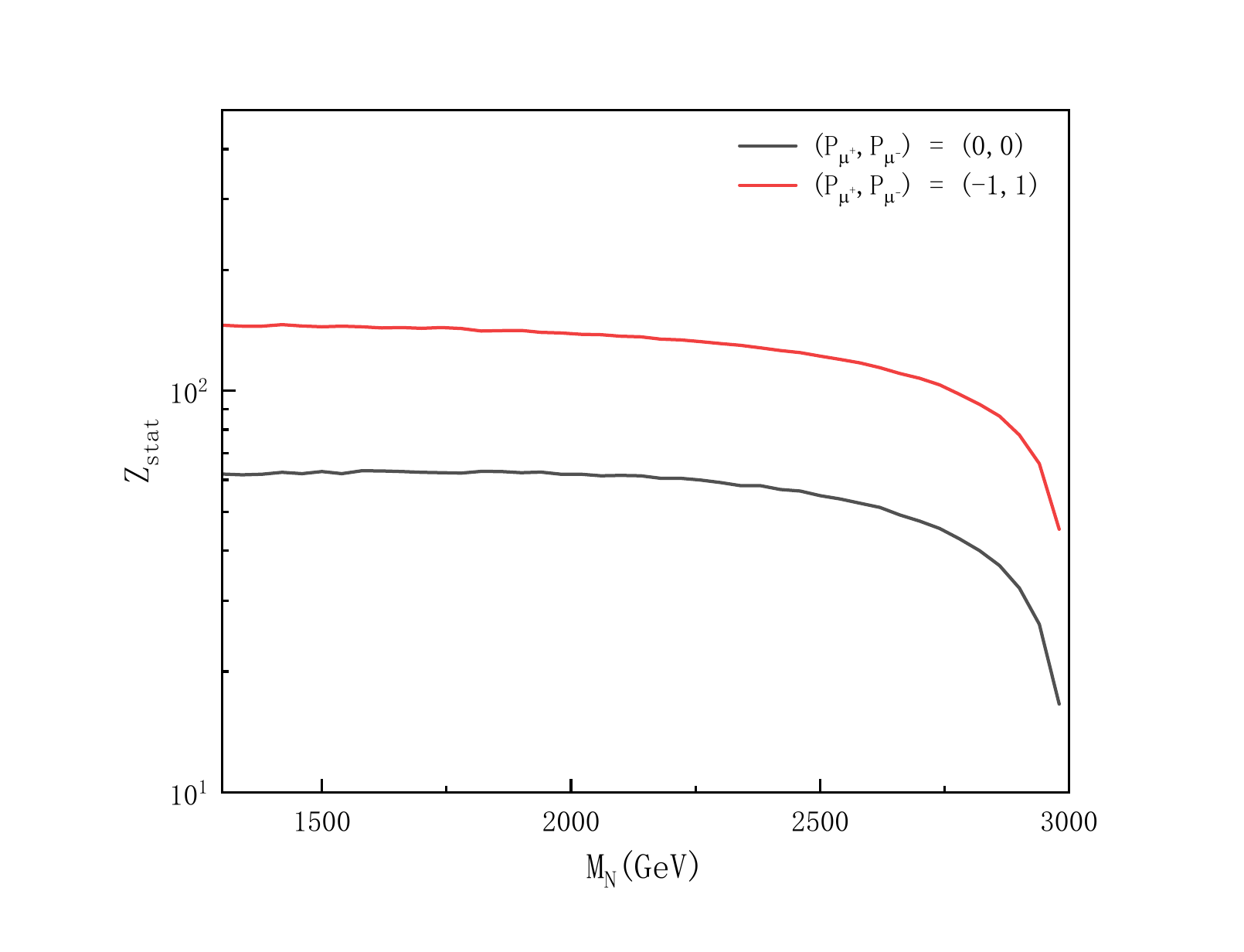}
\caption{The statistics-only significance $Z_{\rm stat}$ for the $4j2\mu$ signal for the unpolarized and polarized beam configurations at the $6$ TeV muon collider with $\mathcal{L}=$ $4$ ab$^{-1}$. }
\label{fig:6}
\end{center}
\end{figure}

The very large expected significances in the polarized configuration arise from the high signal-to-background ratio after the optimized selections. For example, at $M_N
=1300$ GeV, the significance decreases from $62.04$ to $47.57$, $36.55$ and $24.48$ in the unpolarized configuration for $\delta_B=5\%,10\%$ and $20\%$, respectively. In comparison, the corresponding values in the polarized configuration decrease from $145.65$ to $126.86$, $113.88$ and $97.91$. A similar trend is observed for the other benchmark points, with the polarized configuration being less affected because of its higher signal-to-background ratio. All benchmark points remain well above the nominal $5\sigma$ discovery threshold. These results should be interpreted as idealized expected sensitivities within the adopted detector and background assumptions, rather than as complete experimental significance estimates.
The mass dependence of the statistics-only significance is shown in Fig.~\ref{fig:6} for the unpolarized and polarized beam configurations. The polarized configuration consistently provides a larger significance over the investigated mass range. $Z_{\rm stat}$ decreases from about $145.65~(62.04)$ at $M_N=1300$ GeV to $45.22~(16.58)$ at $M_N=2980$ GeV for the polarized (unpolarized) case.

\subsection{The $2\ell2\mu E\mkern-10.5 mu/$ signal}

For the $2\ell2\mu E\mkern-10.5 mu/$ signal, we adopt an analysis strategy similar to that used for the $4j2\mu$ signal. For the polarized configuration $(P_{\mu^+},P_{\mu^-})=(-1,1)$, the signal cross section decreases from $9.71\times 10^{-2}$ fb to $1.70\times 10^{-2}$ fb for VLL masses between $1300$ and $2980$ GeV at the $6$ TeV muon collider. For comparison, the corresponding unpolarized cross section decreases from $5.58\times 10^{-2}$ fb to $9.75\times 10^{-3}$ fb over the same mass range. The main SM background processes considered are:
(I) $\mu^+\mu^-\rightarrow \ell^+\ell^-\mu^+\mu^-\nu_\ell\bar{\nu_\ell}$;
(II) $\mu^+\mu^-\rightarrow W^+W^-W^+W^-$, where all four $W$ bosons decay leptonically, leading to the same charged-lepton content as the signal but with four neutrinos in the final state;
(III) $\mu^+\mu^-\rightarrow \ell^+\ell^-\mu^+\mu^-$. Although this process contains no genuine source of missing transverse momentum, detector effects can generate apparent missing transverse momentum, allowing it to enter the signal region. The processes in (I) and (III) are generated as complete tree-level electroweak final states, including all relevant diagrams and their interference.

\begin{figure}[H]
\begin{center}
\subfigure[]{\includegraphics [scale=0.33] {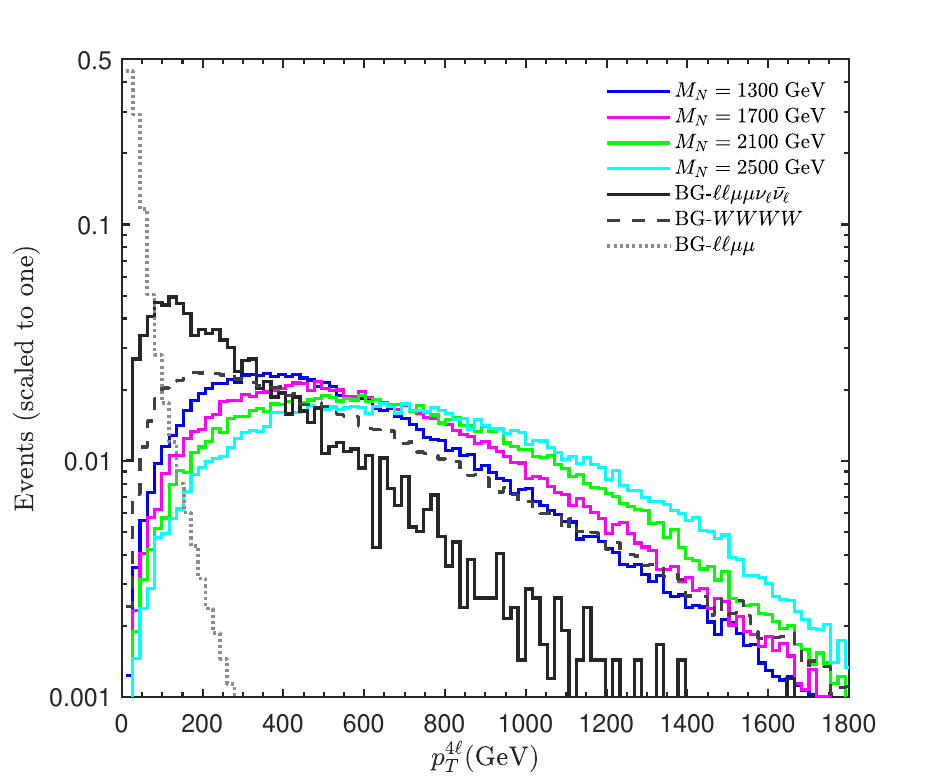}}
\subfigure[]{\includegraphics [scale=0.33] {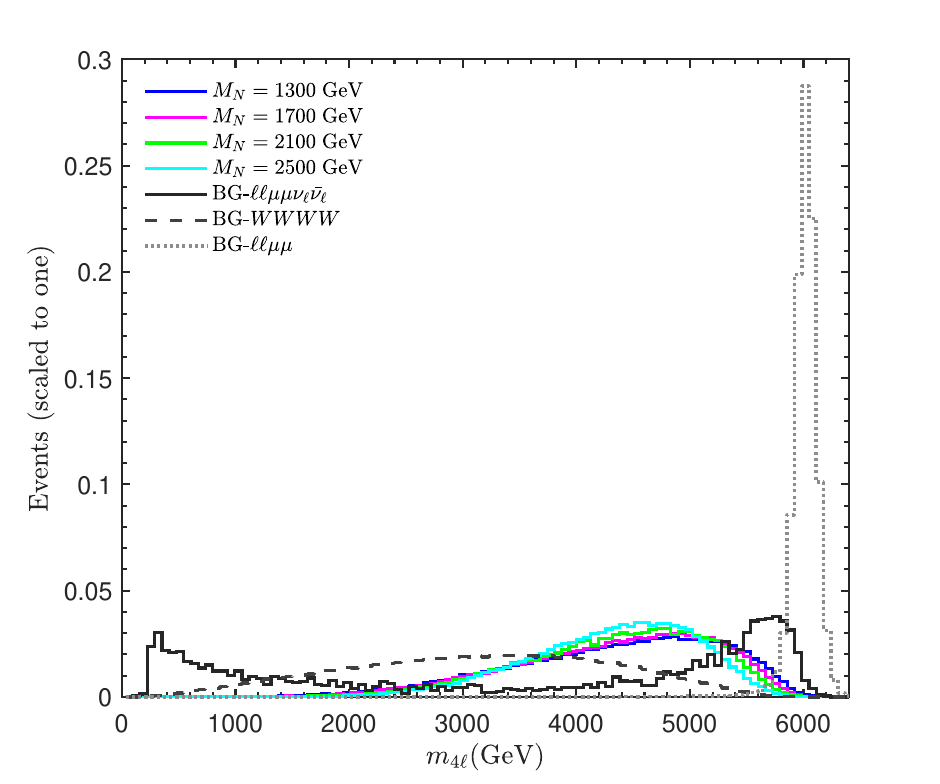}}
\subfigure[]{\includegraphics [scale=0.33] {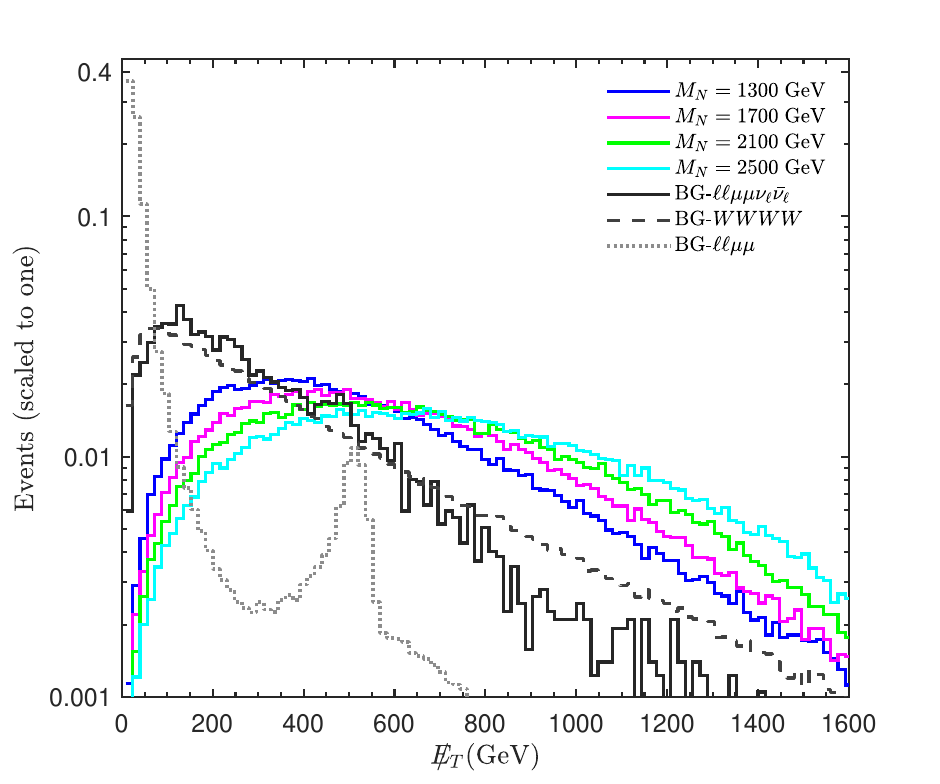}}
\vspace{-0.3cm}
\caption{The normalized distributions of the observables $p_T^{4\ell}$ (a), $m_{4\ell}$ (b) and $E\mkern-10.5 mu/_T$ (c) for the signals of selected neutral VLL-mass benchmark points and the SM backgrounds at the $6$ TeV muon collider with $\mathcal{L}=$ $4$ ab$^{-1}$ for the unpolarized beam configuration $(P_{\mu^+},P_{\mu^-})=(0,0)$.}
\label{fig:7}
\end{center}
\end{figure}

\begin{figure}[H]
\begin{center}
\subfigure[]{\includegraphics [scale=0.33] {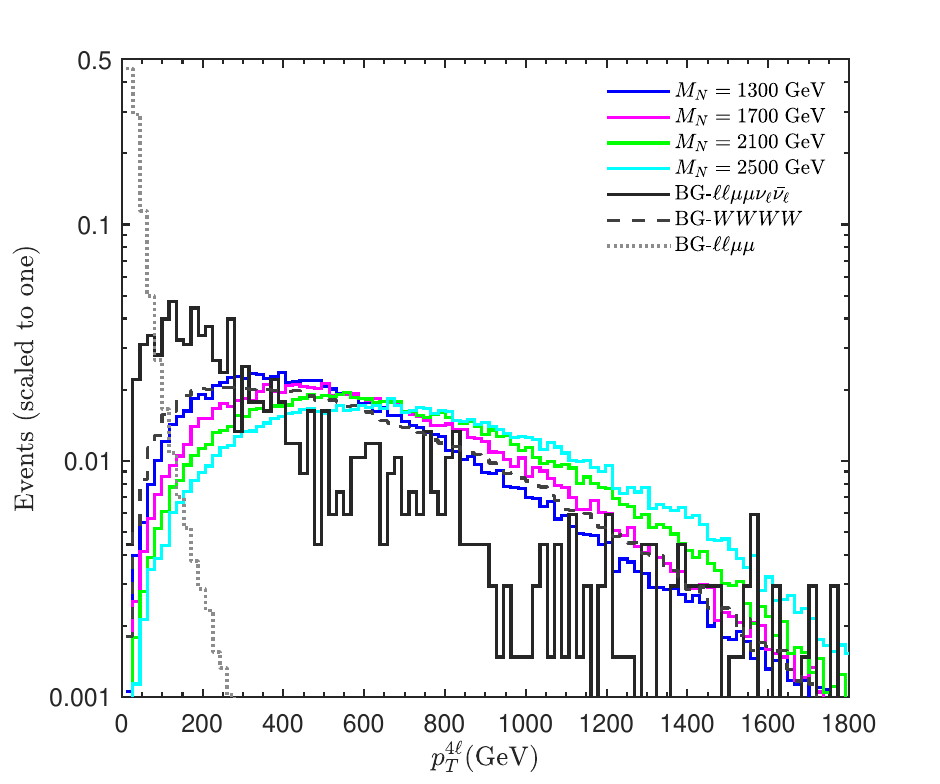}}
\subfigure[]{\includegraphics [scale=0.33] {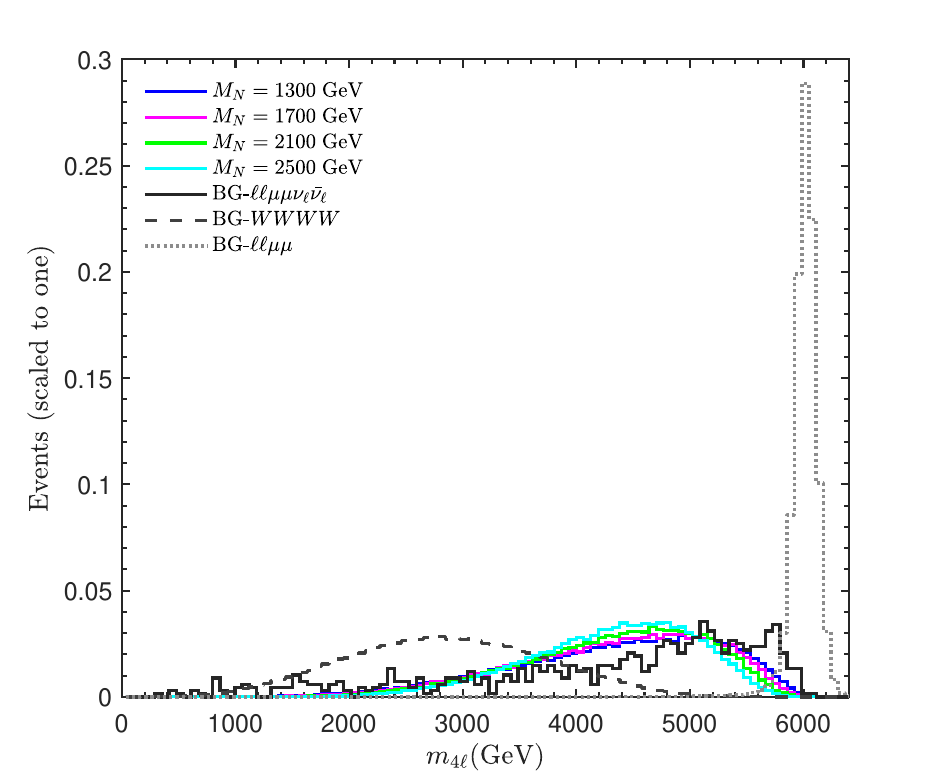}}
\subfigure[]{\includegraphics [scale=0.33] {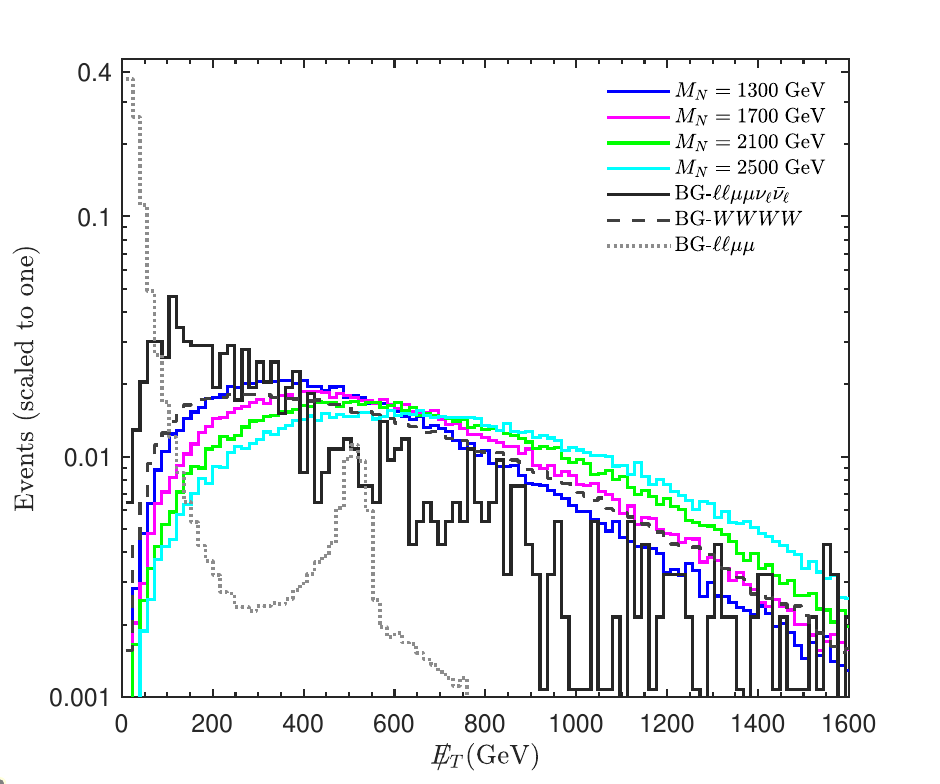}}
\vspace{-0.3cm}
\caption{Same as Fig.~\ref{fig:7} but for the polarized beam configuration $(P_{\mu^+},P_{\mu^-})=(-1,1)$.}
\label{fig:8}
\end{center}
\end{figure}

To further separate the signal from the SM backgrounds, we consider the transverse momentum and invariant mass of the four-lepton system, $p_T^{4\ell}$ and $m_{4\ell}$, together with the missing transverse energy, $E\mkern-10.5 mu/_T$. Since $p_T^{4\ell}$ and $m_{4\ell}$
are constructed from all four reconstructed leptons, there is no need to distinguish the muons produced directly in the $N\rightarrow W\mu$ decay from those originating from the leptonic $W$ decays. 
The normalized distributions of these observables after the basic selections are shown in Figs.~\ref{fig:7} and~\ref{fig:8} for the unpolarized and polarized beam configurations, respectively.

As shown in Figs.~\ref{fig:7}(a) and~\ref{fig:8}(a), the signal has a broad $p_T^{4\ell}$ distribution extending to larger values than the $\ell\ell\mu\mu$ background. This is mainly because the two neutrinos from the leptonic $W$ boson decays carry away transverse momentum, resulting in a larger transverse momentum of the visible four-lepton system. In contrast, the $\ell\ell\mu\mu$ background contains no genuine invisible particles at the matrix-element level and is mainly concentrated at lower $p_T^{4\ell}$. The backgrounds containing neutrinos show broader distributions and partly overlap with the signal. The four-lepton invariant mass $m_{4\ell}$ provides complementary discrimination. For the signal, part of the collision energy is carried away by neutrinos, so the visible four-lepton invariant mass lies below the center-of-mass energy. In contrast, the $\ell\ell\mu\mu$ background is strongly concentrated near $\sqrt{s}=6$ TeV, while the backgrounds containing neutrinos show broader distributions at lower invariant masses, as shown in Figs.~\ref{fig:7}(b) and~\ref{fig:8}(b). This explains why an upper cut on $m_{4\ell}$ is effective, even though the signal originates from a heavy VLL. We also consider $E\mkern-10.5 mu/_T$, as shown in Figs.~\ref{fig:7}(c) and ~\ref{fig:8}(c). The signal contains two neutrinos from the leptonic $W$ boson decays and therefore has a broad distribution extending to relatively large $E\mkern-10.5 mu/_T$. The $\ell\ell\mu\mu$ background is mainly concentrated at low $E\mkern-10.5 mu/_T$, although detector effects generate a residual high-$E\mkern-10.5 mu/_T$ tail. The backgrounds with additional neutrinos also populate larger $E\mkern-10.5 mu/_T$ values.

\begin{table}[H]
\begin{center}
\renewcommand{\arraystretch}{1}
\setlength{\tabcolsep}{18pt}{
\caption{Same as Table~\ref{tab1} but for the $2\ell2\mu E\mkern-10.5 mu/$ signal.}
\label{tab5}
\resizebox{0.9\textwidth}{!}{
\begin{tabular}[c]{c|c|c}
\hline\hline
 & $(P_{\mu^+},P_{\mu^-})=(0,0)$ & $(P_{\mu^+},P_{\mu^-})=(-1,1)$ \\
\hline
\multirow{1}{*}{Basic cuts}
& $p_T^\ell > 10 \:\text{GeV}$,\: $|\eta_\ell| < 2.5$,\: $\Delta R_{\ell\ell} > 0.4$ & $p_T^\ell > 10 \:\text{GeV}$,\: $|\eta_\ell| < 2.5$,\: $\Delta R_{\ell\ell} > 0.4$\\
\hline
Cut 1
& $p_T^{4\ell} > 300 ~\mathrm{GeV}$
& $p_T^{4\ell} > 200 ~\mathrm{GeV}$ \\
Cut 2
& $m_{4\ell} < 5500~\mathrm{GeV}$
& $m_{4\ell} < 5800~\mathrm{GeV}$ \\
Cut 3
& $E\mkern-10.5 mu/_T > 200~\mathrm{GeV}$
& $E\mkern-10.5 mu/_T > 100~\mathrm{GeV}$ \\
\hline\hline
\end{tabular}}}
\end{center}
\end{table}

Since the relative background contributions and detailed kinematic distributions differ between the unpolarized and polarized beam configurations, the selections are optimized separately. The final selection cuts are summarized in Table~\ref{tab5}. The cross sections and corresponding expected numbers of events for the signal benchmark points and the SM backgrounds are presented in Tables~\ref{tab6} and~\ref{tab7} for the unpolarized and polarized beam configurations, respectively. The expected event numbers are evaluated for an integrated luminosity of $\mathcal{L}=4$ ab$^{-1}$.

\begin{table}[H]
\scriptsize
\renewcommand{\arraystretch}{1.3}
\setlength{\tabcolsep}{24pt}
\centering
\caption{Same as Table~\ref{tab2} but for the $2\ell2\mu E\mkern-10.5 mu/$ signal. }
\label{tab6}
\resizebox{0.95\textwidth}{!}{
\begin{tabular}{c|cccc}
\hline\hline
\multirow{2}{*}{Process}
& \multicolumn{4}{c}{cross sections for signal and backgrounds [fb] (expected events)} \\
\cline{2-5}
& Basic cuts
& Cut 1
& Cut 2
& Cut 3 \\
\hline

$M_N=1300~\mathrm{GeV}$
& \makecell{$5.58\times10^{-2}$\\$(223)$}
& \makecell{$2.98\times10^{-2}$\\$(119)$}
& \makecell{$2.91\times10^{-2}$\\$(116)$}
& \makecell{$2.88\times10^{-2}$\\$(115)$} \\

\hline

$M_N=1700~\mathrm{GeV}$
& \makecell{$5.41\times10^{-2}$\\$(216)$}
& \makecell{$3.11\times10^{-2}$\\$(124)$}
& \makecell{$3.05\times10^{-2}$\\$(122)$}
& \makecell{$3.03\times10^{-2}$\\$(121)$} \\

\hline

$M_N=2100~\mathrm{GeV}$
& \makecell{$5.03\times10^{-2}$\\$(201)$}
& \makecell{$3.02\times10^{-2}$\\$(121)$}
& \makecell{$2.98\times10^{-2}$\\$(119)$}
& \makecell{$2.96\times10^{-2}$\\$(118)$} \\

\hline

$M_N=2500~\mathrm{GeV}$
& \makecell{$4.22\times10^{-2}$\\$(168)$}
& \makecell{$2.63\times10^{-2}$\\$(105)$}
& \makecell{$2.62\times10^{-2}$\\$(105)$}
& \makecell{$2.60\times10^{-2}$\\$(104)$} \\

\hline

$BG-\ell\ell\mu\mu\nu_\ell\bar{\nu_\ell}$
& \makecell{$1.01$\\$(4030)$}
& \makecell{$3.05\times10^{-1}$\\$(1221)$}
& \makecell{$2.92\times10^{-1}$\\$(1168)$}
& \makecell{$2.88\times10^{-1}$\\$(1150)$}\\

\hline

$BG-WWWW$
& \makecell{$1.64\times10^{-3}$\\$(6.57)$}
& \makecell{$4.23\times10^{-4}$\\$(1.69)$}
& \makecell{$4.20\times10^{-4}$\\$(1.68)$}
& \makecell{$4.18\times10^{-4}$\\$(1.67)$}\\

\hline

$BG-\ell\ell\mu\mu$
& \makecell{$4.94\times10^{-1}$\\$(1977)$}
& \makecell{$2.06\times10^{-3}$\\$(8.23)$}
& \makecell{$1.45\times10^{-3}$\\$(5.80)$}
& \makecell{$1.19\times10^{-4}$\\$(4.75\times10^{-1})$}\\

\hline\hline
\end{tabular}
}
\end{table}

\begin{table}[H]
\scriptsize
\renewcommand{\arraystretch}{1.3}
\setlength{\tabcolsep}{24pt}
\centering
\caption{Same as Table~\ref{tab3} but for the $2\ell2\mu E\mkern-10.5 mu/$ signal.}
\label{tab7}
\resizebox{0.95\textwidth}{!}{
\begin{tabular}{c|cccc}
\hline\hline
\multirow{2}{*}{Process}
& \multicolumn{4}{c}{cross sections for signal and backgrounds [fb] (expected events)} \\
\cline{2-5}
& Basic cuts
& Cut 1
& Cut 2
& Cut 3 \\
\hline

$M_N=1300~\mathrm{GeV}$
& \makecell{$9.71\times10^{-2}$\\$(388)$}
& \makecell{$5.95\times10^{-2}$\\$(238)$}
& \makecell{$5.93\times10^{-2}$\\$(237)$}
& \makecell{$5.91\times10^{-2}$\\$(236)$} \\

\hline

$M_N=1700~\mathrm{GeV}$
& \makecell{$9.42\times10^{-2}$\\$(376)$}
& \makecell{$6.04\times10^{-2}$\\$(242)$}
& \makecell{$6.03\times10^{-2}$\\$(241)$}
& \makecell{$6.01\times10^{-2}$\\$(240)$} \\

\hline

$M_N=2100~\mathrm{GeV}$
& \makecell{$8.75\times10^{-2}$\\$(349)$}
& \makecell{$5.76\times10^{-2}$\\$(230)$}
& \makecell{$5.75\times10^{-2}$\\$(230)$}
& \makecell{$5.74\times10^{-2}$\\$(230)$} \\

\hline

$M_N=2500~\mathrm{GeV}$
& \makecell{$7.34\times10^{-2}$\\$(293)$}
& \makecell{$4.90\times10^{-2}$\\$(196)$}
& \makecell{$4.90\times10^{-2}$\\$(196)$}
& \makecell{$4.89\times10^{-2}$\\$(196)$} \\

\hline

$BG-\ell\ell\mu\mu\nu_\ell\bar{\nu_\ell}$
& \makecell{$2.44\times10^{-3}$\\$(9.75)$}
& \makecell{$1.15\times10^{-3}$\\$(4.61)$}
& \makecell{$1.14\times10^{-3}$\\$(4.57)$}
& \makecell{$1.14\times10^{-3}$\\$(4.56)$}\\

\hline

$BG-WWWW$
& \makecell{$7.73\times10^{-6}$\\$(3.09\times10^{-2})$}
& \makecell{$4.50\times10^{-6}$\\$(1.80\times10^{-2})$}
& \makecell{$4.50\times10^{-6}$\\$(1.80\times10^{-2})$}
& \makecell{$4.48\times10^{-6}$\\$(1.79\times10^{-2})$}\\

\hline

$BG-\ell\ell\mu\mu$
& \makecell{$5.65\times10^{-1}$\\$(2261)$}
& \makecell{$5.19\times10^{-3}$\\$(21)$}
& \makecell{$3.79\times10^{-3}$\\$(15)$}
& \makecell{$4.70\times10^{-4}$\\$(1.88)$}\\

\hline\hline
\end{tabular}
}
\end{table}

The statistics-only significance is calculated using the same definition as in the $4j2\mu$ analysis. To assess the impact of background systematic uncertainties, we also consider the same fractional background-normalization uncertainties, $\delta_B=5\%,10\%$ and $20\%$. The resulting significances for several benchmark values of $M_N$ are summarized in Table~\ref{tab8}. Taking $M_N=1700$ GeV as an example, the unpolarized significance decreases from $3.50$ in the statistics-only case to $0.52$ for $\delta_B=20\%$, whereas the polarized significance decreases from $36.31$ to $29.80$. The same pattern is observed for the other benchmark points in Table~\ref{tab8}. This shows that the unpolarized results are more sensitive to the residual-background uncertainty, while the higher signal-to-background ratio in the polarized configuration makes the significance more stable.

\begin{table}[H]
\scriptsize
\renewcommand{\arraystretch}{1.3}
\setlength{\tabcolsep}{8pt}
\centering
\caption{Same as Table~\ref{tab4} but for the $2\ell2\mu E\mkern-10.5 mu/$ signal.}
\label{tab8}

\resizebox{0.95\textwidth}{!}{
\begin{tabular}{c|cccc|cccc}
\hline\hline
\multirow{2}{*}{$M_N$ [GeV]}
& \multicolumn{4}{c|}{$(P_{\mu^+},P_{\mu^-})=(0,0)$}
& \multicolumn{4}{c}{$(P_{\mu^+},P_{\mu^-})=(-1,1)$} \\
\cline{2-9}
& $Z_{\rm stat}$
& $Z_{\delta_B=5\%}$
& $Z_{\delta_B=10\%}$
& $Z_{\delta_B=20\%}$
& $Z_{\rm stat}$
& $Z_{\delta_B=5\%}$
& $Z_{\delta_B=10\%}$
& $Z_{\delta_B=20\%}$ \\
\hline

1300
& $3.34$ & $1.70$ & $0.96$ & $0.49$
& $35.90$ & $34.56$ & $32.62$ & $29.48$ \\

\hline

1700
& $3.50$ & $1.79$ & $1.00$ & $0.52$
& $36.31$ & $34.95$ & $32.97$ & $29.80$ \\

\hline

2100
& $3.43$ & $1.75$ & $0.98$ & $0.51$
& $35.21$ & $33.91$ & $32.01$ & $28.93$ \\

\hline

2500
& $3.02$ & $1.54$ & $0.86$ & $0.45$
& $31.64$ & $30.55$ & $28.87$ & $26.08$ \\

\hline\hline
\end{tabular}
}
\end{table}

\begin{figure}[H]
\begin{center}
\centering\includegraphics [scale=0.4] {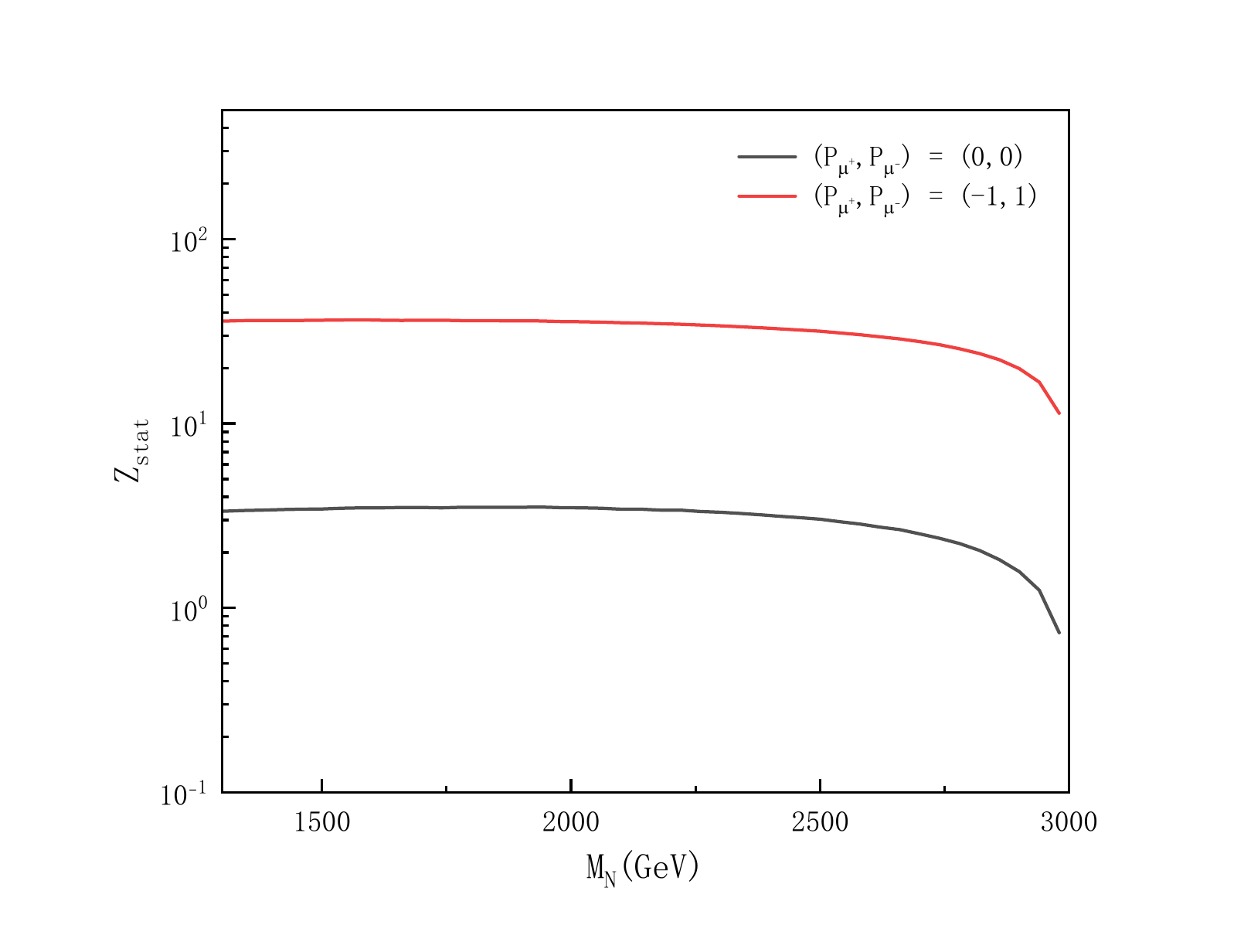}
\caption{Same as Fig.~\ref{fig:6} but for the $2\ell2\mu E\mkern-10.5 mu/$ signal. }
\label{fig:9}
\end{center}
\end{figure}

The dependence of $Z_{\rm stat}$ on $M_N$ over the full mass range is shown in Fig.~\ref{fig:9} for the unpolarized and polarized beam configurations. For the polarized case, $Z_{\rm stat}$ decreases from about $35.90$ at $M_N=1300$ GeV to around $11.36$ at $M_N=2980$ GeV, while the unpolarized significance decreases from $3.34$ to $0.73$ over the same mass range.
The polarized beam configuration provides a much higher sensitivity than the unpolarized case. For the unpolarized configuration, the statistics-only significance remains below the $5\sigma$ discovery level over the considered mass range and decreases rapidly as $M_N$ approaches the kinematic limit. In contrast, the polarized configuration maintains a statistics-only significance above $5\sigma$ throughout the explored mass range. The inclusion of background normalization uncertainties reduces the significance, particularly in the unpolarized case, while the polarized results remain well above the discovery level for the benchmark masses shown in Table~\ref{tab8} even for $\delta_B=20\%$.

Overall, the $4j2\mu$ signal provides a higher sensitivity than the $2\ell2\mu E\mkern-10.5 mu/$ signal over most of the considered VLL-mass range under the assumptions of the present fast-simulation analysis. This mainly reflects the larger signal rate and the more effective background suppression in the $4j2\mu$ signal. Beam polarization increases the expected sensitivity in both signals. In the $2\ell2\mu E\mkern-10.5 mu/$ signal, the larger residual backgrounds lead to a stronger reduction of the  significance once background normalization uncertainties are included. The two signals therefore provide complementary probes of neutral VLL pair production at the muon collider.

%%%%%%%%%%%%%%%%%%%%%%%%%%%%%%%%%%%%%%%%%%%%%%%%%%%%%%%%%%%%%%%%%%%%%%%%%%%%%%%%%%%%%%%%%%%%%%%%%%%%%%%%%%%%%%%%%%%%%%%%%%%%%%%%%%%%
\section{Conclusions and discussions}

Vector-like leptons (VLLs) are well-motivated candidates for physics beyond the SM and may help address several open questions in particle physics. They appear naturally in many BSM scenarios and are therefore promising targets for experimental searches. Due to their distinctive properties, VLLs can give rise to rich phenomenology, making them relevant for both current and future experimental studies.

In this work, we consider a doublet VLL model in which the VLLs mix with second-generation SM leptons and study the pair production of the neutral VLL $N$ at the $6$ TeV muon collider with $\mathcal{L}=$ $4$ ab$^{-1}$.
Considering both the unpolarized and polarized beam configurations, we investigate the $4j2\mu$ and $2\ell2\mu E\mkern-10.5 mu/$ signals arising from the decay $N\rightarrow W^+\mu^-$. Under the detector, polarization, luminosity and background assumptions adopted in this analysis, the $6$ TeV muon collider could provide strong sensitivity to neutral VLLs in the mass range of $1300-2980$ GeV through $4j2\mu$ and $2\ell2\mu E\mkern-10.5 mu/$ signals. For the idealized fully polarized configuration $(P_{\mu^+},P_{\mu^-})=(-1,1)$, the statistics-only significance decreases from $145.65$ to $45.22$ for the $4j2\mu$ signal and from $35.90$ to $11.36$ for the $2\ell2\mu E\mkern-10.5 mu/$ signal over this mass range. These results indicate that the $4j2\mu$ signal provides higher sensitivity than the $2\ell2\mu E\mkern-10.5 mu/$ signal. These sensitivities should therefore be interpreted as idealized fast-simulation projections under the adopted assumptions, rather than as complete experimental estimates.

The fully polarized configuration considered in this work is an idealized benchmark used to illustrate the impact of beam polarization. Since achieving full polarization at a high-energy muon collider would be experimentally challenging, a realistic beam would more likely involve partial polarization. In this case, the improvement in sensitivity would generally be smaller than that obtained with full polarization and the significance would approach the unpolarized result as the degree of polarization decreases. Its precise value would depend on the achievable polarization and the corresponding changes in both the signal and background rates. The unpolarized results are therefore also presented as a reference. In addition, background-normalization uncertainties reduce the expected significance, especially when the residual background is large relative to the signal. Their impact is estimated here using representative values of the background uncertainty. A complete assessment of the expected sensitivity would require a dedicated study that includes realistic partial beam polarization, detector-related systematic uncertainties and beam-induced backgrounds.

It should be emphasized that this work focuses on the neutral $N\bar{N}$ pair-production channel rather than on the complete phenomenology of the VLL doublet model. The charged partner $E$, which is approximately mass-degenerate with $N$, can also be pair produced at the muon collider and may provide complementary information on the doublet scenario.
Therefore, the results presented here should be interpreted as the sensitivity of the selected neutral $N\bar{N}$ channel rather than the combined sensitivity of the complete VLL doublet model, for which a dedicated study of the charged partner channel is required. Nevertheless, the neutral VLL pair-production channel studied here may serve as a useful reference for future studies of vector-like leptons at high-energy muon colliders.

%%%%%%%%%%%%%%%%%%%%%%%%%%%%%%%%%%%%%%%%%%%%%%%%%%%%%%%%%%%%%%%%%%%%%%%%%%%%%%%%%%%%%%%%%%%%%%%%%%%%%%%%%%%%%%%%%%%%%%%%%%%%%%%%%%%%
\section*{ACKNOWLEDGMENT}

This work was partially supported by the National Natural Science Foundation of China under Grant No. 12575106 and the Cultivation Fund of Liaoning Normal University for Excellent Doctoral Dissertations~(No. YJSYB202501).

%%%%%%%%%%%%%%%%%%%%%%%%%%%%%%%%%%%%%%%%%%%%%%%%%%%%%%%%%%%%%%%


\begin{thebibliography}{101}
\expandafter\ifx\csname natexlab\endcsname\relax\def\natexlab#1{#1}\fi
\expandafter\ifx\csname bibnamefont\endcsname\relax
  \def\bibnamefont#1{#1}\fi
\expandafter\ifx\csname bibfnamefont\endcsname\relax
  \def\bibfnamefont#1{#1}\fi
\expandafter\ifx\csname citenamefont\endcsname\relax
  \def\citenamefont#1{#1}\fi
\expandafter\ifx\csname url\endcsname\relax
  \def\url#1{\texttt{#1}}\fi
\expandafter\ifx\csname urlprefix\endcsname\relax\def\urlprefix{URL }\fi
\providecommand{\bibinfo}[2]{#2}
\providecommand{\eprint}[2][]{\url{#2}}

\bibitem[{\citenamefont{Salam}(1968)}]{Salam:1968rm}
\bibinfo{author}{\bibfnamefont{A.}~\bibnamefont{Salam}},
  \bibinfo{journal}{Conf. Proc. C} \textbf{\bibinfo{volume}{680519}},
  \bibinfo{pages}{367} (\bibinfo{year}{1968}).

\bibitem[{\citenamefont{Weinberg}(1967)}]{Weinberg:1967tq}
\bibinfo{author}{\bibfnamefont{S.}~\bibnamefont{Weinberg}},
  \bibinfo{journal}{Phys. Rev. Lett.} \textbf{\bibinfo{volume}{19}},
  \bibinfo{pages}{1264} (\bibinfo{year}{1967}).

\bibitem[{\citenamefont{Glashow}(1961)}]{Glashow:1961tr}
\bibinfo{author}{\bibfnamefont{S.~L.} \bibnamefont{Glashow}},
  \bibinfo{journal}{Nucl. Phys.} \textbf{\bibinfo{volume}{22}},
  \bibinfo{pages}{579} (\bibinfo{year}{1961}).

\bibitem[{\citenamefont{Feng}(2013)}]{Feng:2013pwa}
\bibinfo{author}{\bibfnamefont{J.~L.} \bibnamefont{Feng}},
  \bibinfo{journal}{Ann. Rev. Nucl. Part. Sci.} \textbf{\bibinfo{volume}{63}},
  \bibinfo{pages}{351} (\bibinfo{year}{2013}), \eprint{1302.6587}.

\bibitem[{\citenamefont{Duffy and van Bibber}(2009)}]{Duffy:2009ig}
\bibinfo{author}{\bibfnamefont{L.~D.} \bibnamefont{Duffy}} \bibnamefont{and}
  \bibinfo{author}{\bibfnamefont{K.}~\bibnamefont{van Bibber}},
  \bibinfo{journal}{New J. Phys.} \textbf{\bibinfo{volume}{11}},
  \bibinfo{pages}{105008} (\bibinfo{year}{2009}), \eprint{0904.3346}.

\bibitem[{\citenamefont{Chadha-Day et~al.}(2022)\citenamefont{Chadha-Day,
  Ellis, and Marsh}}]{Chadha-Day:2021szb}
\bibinfo{author}{\bibfnamefont{F.}~\bibnamefont{Chadha-Day}},
  \bibinfo{author}{\bibfnamefont{J.}~\bibnamefont{Ellis}}, \bibnamefont{and}
  \bibinfo{author}{\bibfnamefont{D.~J.~E.} \bibnamefont{Marsh}},
  \bibinfo{journal}{Sci. Adv.} \textbf{\bibinfo{volume}{8}},
  \bibinfo{pages}{abj3618} (\bibinfo{year}{2022}), \eprint{2105.01406}.

\bibitem[{\citenamefont{Gonzalez-Garcia and
  Maltoni}(2008)}]{Gonzalez-Garcia:2007dlo}
\bibinfo{author}{\bibfnamefont{M.~C.} \bibnamefont{Gonzalez-Garcia}}
  \bibnamefont{and} \bibinfo{author}{\bibfnamefont{M.}~\bibnamefont{Maltoni}},
  \bibinfo{journal}{Phys. Rept.} \textbf{\bibinfo{volume}{460}},
  \bibinfo{pages}{1} (\bibinfo{year}{2008}), \eprint{0704.1800}.

\bibitem[{\citenamefont{Hiller et~al.}(2020)\citenamefont{Hiller,
  Hormigos-Feliu, Litim, and Steudtner}}]{Hiller:2019mou}
\bibinfo{author}{\bibfnamefont{G.}~\bibnamefont{Hiller}},
  \bibinfo{author}{\bibfnamefont{C.}~\bibnamefont{Hormigos-Feliu}},
  \bibinfo{author}{\bibfnamefont{D.~F.} \bibnamefont{Litim}}, \bibnamefont{and}
  \bibinfo{author}{\bibfnamefont{T.}~\bibnamefont{Steudtner}},
  \bibinfo{journal}{Phys. Rev. D} \textbf{\bibinfo{volume}{102}},
  \bibinfo{pages}{071901} (\bibinfo{year}{2020}), \eprint{1910.14062}.

\bibitem[{\citenamefont{Frank and Saha}(2020)}]{Frank:2020smf}
\bibinfo{author}{\bibfnamefont{M.}~\bibnamefont{Frank}} \bibnamefont{and}
  \bibinfo{author}{\bibfnamefont{I.}~\bibnamefont{Saha}},
  \bibinfo{journal}{Phys. Rev. D} \textbf{\bibinfo{volume}{102}},
  \bibinfo{pages}{115034} (\bibinfo{year}{2020}), \eprint{2008.11909}.

\bibitem[{\citenamefont{Dermisek et~al.}(2021)\citenamefont{Dermisek, Hermanek,
  and McGinnis}}]{Dermisek:2021ajd}
\bibinfo{author}{\bibfnamefont{R.}~\bibnamefont{Dermisek}},
  \bibinfo{author}{\bibfnamefont{K.}~\bibnamefont{Hermanek}}, \bibnamefont{and}
  \bibinfo{author}{\bibfnamefont{N.}~\bibnamefont{McGinnis}},
  \bibinfo{journal}{Phys. Rev. D} \textbf{\bibinfo{volume}{104}},
  \bibinfo{pages}{055033} (\bibinfo{year}{2021}), \eprint{2103.05645}.

\bibitem[{\citenamefont{Brune et~al.}(2024)\citenamefont{Brune, Kephart, and
  P{\"a}s}}]{Brune:2022rlo}
\bibinfo{author}{\bibfnamefont{T.}~\bibnamefont{Brune}},
  \bibinfo{author}{\bibfnamefont{T.~W.} \bibnamefont{Kephart}},
  \bibnamefont{and} \bibinfo{author}{\bibfnamefont{H.}~\bibnamefont{P{\"a}s}},
  \bibinfo{journal}{Eur. Phys. J. C} \textbf{\bibinfo{volume}{84}},
  \bibinfo{pages}{1254} (\bibinfo{year}{2024}), \eprint{2205.05566}.

\bibitem[{\citenamefont{Guedes and Olgoso}(2022)}]{Guedes:2022cfy}
\bibinfo{author}{\bibfnamefont{G.}~\bibnamefont{Guedes}} \bibnamefont{and}
  \bibinfo{author}{\bibfnamefont{P.}~\bibnamefont{Olgoso}},
  \bibinfo{journal}{JHEP} \textbf{\bibinfo{volume}{09}}, \bibinfo{pages}{181}
  (\bibinfo{year}{2022}), \eprint{2205.04480}.

\bibitem[{\citenamefont{M{\"o}hling et~al.}(2024)\citenamefont{M{\"o}hling,
  St{\"o}ckinger, and St{\"o}ckinger-Kim}}]{Mohling:2024qvk}
\bibinfo{author}{\bibfnamefont{K.}~\bibnamefont{M{\"o}hling}},
  \bibinfo{author}{\bibfnamefont{D.}~\bibnamefont{St{\"o}ckinger}},
  \bibnamefont{and}
  \bibinfo{author}{\bibfnamefont{H.}~\bibnamefont{St{\"o}ckinger-Kim}},
  \bibinfo{journal}{JHEP} \textbf{\bibinfo{volume}{10}}, \bibinfo{pages}{170}
  (\bibinfo{year}{2024}), \eprint{2407.09421}.

\bibitem[{\citenamefont{Daberstiel et~al.}(2026)\citenamefont{Daberstiel,
  M{\"o}hling, St{\"o}ckinger, and St{\"o}ckinger-Kim}}]{Daberstiel:2026wxe}
\bibinfo{author}{\bibfnamefont{G.}~\bibnamefont{Daberstiel}},
  \bibinfo{author}{\bibfnamefont{K.}~\bibnamefont{M{\"o}hling}},
  \bibinfo{author}{\bibfnamefont{D.}~\bibnamefont{St{\"o}ckinger}},
  \bibnamefont{and}
  \bibinfo{author}{\bibfnamefont{H.}~\bibnamefont{St{\"o}ckinger-Kim}},
  \bibinfo{journal}{JHEP} \textbf{\bibinfo{volume}{07}}, \bibinfo{pages}{259}
  (\bibinfo{year}{2026}), \eprint{2603.21414}.

\bibitem[{\citenamefont{He}(2023)}]{He:2022zjz}
\bibinfo{author}{\bibfnamefont{S.-P.} \bibnamefont{He}},
  \bibinfo{journal}{Chin. Phys. C} \textbf{\bibinfo{volume}{47}},
  \bibinfo{pages}{043102} (\bibinfo{year}{2023}), \eprint{2205.02088}.

\bibitem[{\citenamefont{Kawamura and Raby}(2022)}]{Kawamura:2022fhm}
\bibinfo{author}{\bibfnamefont{J.}~\bibnamefont{Kawamura}} \bibnamefont{and}
  \bibinfo{author}{\bibfnamefont{S.}~\bibnamefont{Raby}},
  \bibinfo{journal}{Phys. Rev. D} \textbf{\bibinfo{volume}{106}},
  \bibinfo{pages}{035009} (\bibinfo{year}{2022}), \eprint{2205.10480}.

\bibitem[{\citenamefont{Belfatto et~al.}(2020)\citenamefont{Belfatto, Beradze,
  and Berezhiani}}]{Belfatto:2019swo}
\bibinfo{author}{\bibfnamefont{B.}~\bibnamefont{Belfatto}},
  \bibinfo{author}{\bibfnamefont{R.}~\bibnamefont{Beradze}}, \bibnamefont{and}
  \bibinfo{author}{\bibfnamefont{Z.}~\bibnamefont{Berezhiani}},
  \bibinfo{journal}{Eur. Phys. J. C} \textbf{\bibinfo{volume}{80}},
  \bibinfo{pages}{149} (\bibinfo{year}{2020}), \eprint{1906.02714}.

\bibitem[{\citenamefont{Grossman et~al.}(2020)\citenamefont{Grossman, Passemar,
  and Schacht}}]{Grossman:2019bzp}
\bibinfo{author}{\bibfnamefont{Y.}~\bibnamefont{Grossman}},
  \bibinfo{author}{\bibfnamefont{E.}~\bibnamefont{Passemar}}, \bibnamefont{and}
  \bibinfo{author}{\bibfnamefont{S.}~\bibnamefont{Schacht}},
  \bibinfo{journal}{JHEP} \textbf{\bibinfo{volume}{07}}, \bibinfo{pages}{068}
  (\bibinfo{year}{2020}), \eprint{1911.07821}.

\bibitem[{\citenamefont{Crivellin et~al.}(2020)\citenamefont{Crivellin, Kirk,
  Manzari, and Montull}}]{Crivellin:2020ebi}
\bibinfo{author}{\bibfnamefont{A.}~\bibnamefont{Crivellin}},
  \bibinfo{author}{\bibfnamefont{F.}~\bibnamefont{Kirk}},
  \bibinfo{author}{\bibfnamefont{C.~A.} \bibnamefont{Manzari}},
  \bibnamefont{and} \bibinfo{author}{\bibfnamefont{M.}~\bibnamefont{Montull}},
  \bibinfo{journal}{JHEP} \textbf{\bibinfo{volume}{12}}, \bibinfo{pages}{166}
  (\bibinfo{year}{2020}), \eprint{2008.01113}.

\bibitem[{\citenamefont{Dagli et~al.}(2026)\citenamefont{Dagli, Sultansoy, and
  Toy}}]{Dagli:2026lus}
\bibinfo{author}{\bibfnamefont{B.}~\bibnamefont{Dagli}},
  \bibinfo{author}{\bibfnamefont{S.}~\bibnamefont{Sultansoy}},
  \bibnamefont{and} \bibinfo{author}{\bibfnamefont{I.}~\bibnamefont{Toy}}
  (\bibinfo{year}{2026}), \eprint{2603.05104}.

\bibitem[{\citenamefont{Martin}(2010)}]{Martin:2009bg}
\bibinfo{author}{\bibfnamefont{S.~P.} \bibnamefont{Martin}},
  \bibinfo{journal}{Phys. Rev. D} \textbf{\bibinfo{volume}{81}},
  \bibinfo{pages}{035004} (\bibinfo{year}{2010}), \eprint{0910.2732}.

\bibitem[{\citenamefont{Zheng}(2020)}]{Zheng:2019kqu}
\bibinfo{author}{\bibfnamefont{S.}~\bibnamefont{Zheng}}, \bibinfo{journal}{Eur.
  Phys. J. C} \textbf{\bibinfo{volume}{80}}, \bibinfo{pages}{273}
  (\bibinfo{year}{2020}), \eprint{1904.10145}.

\bibitem[{\citenamefont{Kong et~al.}(2010)\citenamefont{Kong, Park, and
  Rizzo}}]{Kong:2010qd}
\bibinfo{author}{\bibfnamefont{K.}~\bibnamefont{Kong}},
  \bibinfo{author}{\bibfnamefont{S.~C.} \bibnamefont{Park}}, \bibnamefont{and}
  \bibinfo{author}{\bibfnamefont{T.~G.} \bibnamefont{Rizzo}},
  \bibinfo{journal}{JHEP} \textbf{\bibinfo{volume}{07}}, \bibinfo{pages}{059}
  (\bibinfo{year}{2010}), \eprint{1004.4635}.

\bibitem[{\citenamefont{Endo et~al.}(2011)\citenamefont{Endo, Hamaguchi,
  Iwamoto, and Yokozaki}}]{Endo:2011mc}
\bibinfo{author}{\bibfnamefont{M.}~\bibnamefont{Endo}},
  \bibinfo{author}{\bibfnamefont{K.}~\bibnamefont{Hamaguchi}},
  \bibinfo{author}{\bibfnamefont{S.}~\bibnamefont{Iwamoto}}, \bibnamefont{and}
  \bibinfo{author}{\bibfnamefont{N.}~\bibnamefont{Yokozaki}},
  \bibinfo{journal}{Phys. Rev. D} \textbf{\bibinfo{volume}{84}},
  \bibinfo{pages}{075017} (\bibinfo{year}{2011}), \eprint{1108.3071}.

\bibitem[{\citenamefont{Araz et~al.}(2018)\citenamefont{Araz, Banerjee, Frank,
  Fuks, and Goudelis}}]{Araz:2018uyi}
\bibinfo{author}{\bibfnamefont{J.~Y.} \bibnamefont{Araz}},
  \bibinfo{author}{\bibfnamefont{S.}~\bibnamefont{Banerjee}},
  \bibinfo{author}{\bibfnamefont{M.}~\bibnamefont{Frank}},
  \bibinfo{author}{\bibfnamefont{B.}~\bibnamefont{Fuks}}, \bibnamefont{and}
  \bibinfo{author}{\bibfnamefont{A.}~\bibnamefont{Goudelis}},
  \bibinfo{journal}{Phys. Rev. D} \textbf{\bibinfo{volume}{98}},
  \bibinfo{pages}{115009} (\bibinfo{year}{2018}), \eprint{1810.07224}.

\bibitem[{\citenamefont{Anastasiou et~al.}(2009)\citenamefont{Anastasiou,
  Furlan, and Santiago}}]{Anastasiou:2009rv}
\bibinfo{author}{\bibfnamefont{C.}~\bibnamefont{Anastasiou}},
  \bibinfo{author}{\bibfnamefont{E.}~\bibnamefont{Furlan}}, \bibnamefont{and}
  \bibinfo{author}{\bibfnamefont{J.}~\bibnamefont{Santiago}},
  \bibinfo{journal}{Phys. Rev. D} \textbf{\bibinfo{volume}{79}},
  \bibinfo{pages}{075003} (\bibinfo{year}{2009}), \eprint{0901.2117}.

\bibitem[{\citenamefont{Gillioz et~al.}(2012)\citenamefont{Gillioz, Grober,
  Grojean, Muhlleitner, and Salvioni}}]{Gillioz:2012se}
\bibinfo{author}{\bibfnamefont{M.}~\bibnamefont{Gillioz}},
  \bibinfo{author}{\bibfnamefont{R.}~\bibnamefont{Grober}},
  \bibinfo{author}{\bibfnamefont{C.}~\bibnamefont{Grojean}},
  \bibinfo{author}{\bibfnamefont{M.}~\bibnamefont{Muhlleitner}},
  \bibnamefont{and} \bibinfo{author}{\bibfnamefont{E.}~\bibnamefont{Salvioni}},
  \bibinfo{journal}{JHEP} \textbf{\bibinfo{volume}{10}}, \bibinfo{pages}{004}
  (\bibinfo{year}{2012}), \eprint{1206.7120}.

\bibitem[{\citenamefont{Agashe et~al.}(2007)\citenamefont{Agashe, Perez, and
  Soni}}]{Agashe:2006wa}
\bibinfo{author}{\bibfnamefont{K.}~\bibnamefont{Agashe}},
  \bibinfo{author}{\bibfnamefont{G.}~\bibnamefont{Perez}}, \bibnamefont{and}
  \bibinfo{author}{\bibfnamefont{A.}~\bibnamefont{Soni}},
  \bibinfo{journal}{Phys. Rev. D} \textbf{\bibinfo{volume}{75}},
  \bibinfo{pages}{015002} (\bibinfo{year}{2007}), \eprint{hep-ph/0606293}.

\bibitem[{\citenamefont{Huang et~al.}(2012)\citenamefont{Huang, Kong, and
  Park}}]{Huang:2012kz}
\bibinfo{author}{\bibfnamefont{G.-Y.} \bibnamefont{Huang}},
  \bibinfo{author}{\bibfnamefont{K.}~\bibnamefont{Kong}}, \bibnamefont{and}
  \bibinfo{author}{\bibfnamefont{S.~C.} \bibnamefont{Park}},
  \bibinfo{journal}{JHEP} \textbf{\bibinfo{volume}{06}}, \bibinfo{pages}{099}
  (\bibinfo{year}{2012}), \eprint{1204.0522}.

\bibitem[{\citenamefont{Tsai et~al.}(1972)\citenamefont{Tsai, Deraad, and
  Milton}}]{Tsai:1972sg}
\bibinfo{author}{\bibfnamefont{W.-Y.} \bibnamefont{Tsai}},
  \bibinfo{author}{\bibfnamefont{L.~L.} \bibnamefont{Deraad}},
  \bibnamefont{and} \bibinfo{author}{\bibfnamefont{K.~A.}
  \bibnamefont{Milton}}, \bibinfo{journal}{Phys. Rev. D}
  \textbf{\bibinfo{volume}{6}}, \bibinfo{pages}{1428} (\bibinfo{year}{1972}),
  \bibinfo{note}{[Erratum: Phys.Rev.D 11, 703--703 (1975)]}.

\bibitem[{\citenamefont{Mohapatra and Pati}(1975)}]{Mohapatra:1974gc}
\bibinfo{author}{\bibfnamefont{R.~N.} \bibnamefont{Mohapatra}}
  \bibnamefont{and} \bibinfo{author}{\bibfnamefont{J.~C.} \bibnamefont{Pati}},
  \bibinfo{journal}{Phys. Rev. D} \textbf{\bibinfo{volume}{11}},
  \bibinfo{pages}{2558} (\bibinfo{year}{1975}).

\bibitem[{\citenamefont{Senjanovic and Mohapatra}(1975)}]{Senjanovic:1975rk}
\bibinfo{author}{\bibfnamefont{G.}~\bibnamefont{Senjanovic}} \bibnamefont{and}
  \bibinfo{author}{\bibfnamefont{R.~N.} \bibnamefont{Mohapatra}},
  \bibinfo{journal}{Phys. Rev. D} \textbf{\bibinfo{volume}{12}},
  \bibinfo{pages}{1502} (\bibinfo{year}{1975}).

\bibitem[{\citenamefont{Mohapatra et~al.}(1978)\citenamefont{Mohapatra, Paige,
  and Sidhu}}]{Mohapatra:1977mj}
\bibinfo{author}{\bibfnamefont{R.~N.} \bibnamefont{Mohapatra}},
  \bibinfo{author}{\bibfnamefont{F.~E.} \bibnamefont{Paige}}, \bibnamefont{and}
  \bibinfo{author}{\bibfnamefont{D.~P.} \bibnamefont{Sidhu}},
  \bibinfo{journal}{Phys. Rev. D} \textbf{\bibinfo{volume}{17}},
  \bibinfo{pages}{2462} (\bibinfo{year}{1978}).

\bibitem[{\citenamefont{Nevzorov}(2013)}]{Nevzorov:2012hs}
\bibinfo{author}{\bibfnamefont{R.}~\bibnamefont{Nevzorov}},
  \bibinfo{journal}{Phys. Rev. D} \textbf{\bibinfo{volume}{87}},
  \bibinfo{pages}{015029} (\bibinfo{year}{2013}), \eprint{1205.5967}.

\bibitem[{\citenamefont{Dorsner et~al.}(2014)\citenamefont{Dorsner, Fajfer, and
  Mustac}}]{Dorsner:2014wva}
\bibinfo{author}{\bibfnamefont{I.}~\bibnamefont{Dorsner}},
  \bibinfo{author}{\bibfnamefont{S.}~\bibnamefont{Fajfer}}, \bibnamefont{and}
  \bibinfo{author}{\bibfnamefont{I.}~\bibnamefont{Mustac}},
  \bibinfo{journal}{Phys. Rev. D} \textbf{\bibinfo{volume}{89}},
  \bibinfo{pages}{115004} (\bibinfo{year}{2014}), \eprint{1401.6870}.

\bibitem[{\citenamefont{Joglekar and Rosner}(2017)}]{Joglekar:2016yap}
\bibinfo{author}{\bibfnamefont{A.}~\bibnamefont{Joglekar}} \bibnamefont{and}
  \bibinfo{author}{\bibfnamefont{J.~L.} \bibnamefont{Rosner}},
  \bibinfo{journal}{Phys. Rev. D} \textbf{\bibinfo{volume}{96}},
  \bibinfo{pages}{015026} (\bibinfo{year}{2017}), \eprint{1607.06900}.

\bibitem[{\citenamefont{Hayrapetyan et~al.}(2025)}]{CMS:2024bni}
\bibinfo{author}{\bibfnamefont{A.}~\bibnamefont{Hayrapetyan}}
  \bibnamefont{et~al.} (\bibinfo{collaboration}{CMS}), \bibinfo{journal}{Phys.
  Rept.} \textbf{\bibinfo{volume}{1115}}, \bibinfo{pages}{570}
  (\bibinfo{year}{2025}), \eprint{2405.17605}.

\bibitem[{\citenamefont{del Aguila et~al.}(2008)\citenamefont{del Aguila,
  de~Blas, and Perez-Victoria}}]{delAguila:2008pw}
\bibinfo{author}{\bibfnamefont{F.}~\bibnamefont{del Aguila}},
  \bibinfo{author}{\bibfnamefont{J.}~\bibnamefont{de~Blas}}, \bibnamefont{and}
  \bibinfo{author}{\bibfnamefont{M.}~\bibnamefont{Perez-Victoria}},
  \bibinfo{journal}{Phys. Rev. D} \textbf{\bibinfo{volume}{78}},
  \bibinfo{pages}{013010} (\bibinfo{year}{2008}), \eprint{0803.4008}.

\bibitem[{\citenamefont{Ishiwata and Wise}(2013)}]{Ishiwata:2013gma}
\bibinfo{author}{\bibfnamefont{K.}~\bibnamefont{Ishiwata}} \bibnamefont{and}
  \bibinfo{author}{\bibfnamefont{M.~B.} \bibnamefont{Wise}},
  \bibinfo{journal}{Phys. Rev. D} \textbf{\bibinfo{volume}{88}},
  \bibinfo{pages}{055009} (\bibinfo{year}{2013}), \eprint{1307.1112}.

\bibitem[{\citenamefont{Thomas and Wells}(1998)}]{Thomas:1998wy}
\bibinfo{author}{\bibfnamefont{S.~D.} \bibnamefont{Thomas}} \bibnamefont{and}
  \bibinfo{author}{\bibfnamefont{J.~D.} \bibnamefont{Wells}},
  \bibinfo{journal}{Phys. Rev. Lett.} \textbf{\bibinfo{volume}{81}},
  \bibinfo{pages}{34} (\bibinfo{year}{1998}), \eprint{hep-ph/9804359}.

\bibitem[{\citenamefont{Achard et~al.}(2001)}]{L3:2001xsz}
\bibinfo{author}{\bibfnamefont{P.}~\bibnamefont{Achard}} \bibnamefont{et~al.}
  (\bibinfo{collaboration}{L3}), \bibinfo{journal}{Phys. Lett. B}
  \textbf{\bibinfo{volume}{517}}, \bibinfo{pages}{75} (\bibinfo{year}{2001}),
  \eprint{hep-ex/0107015}.

\bibitem[{\citenamefont{Tumasyan et~al.}(2022)}]{CMS:2022nty}
\bibinfo{author}{\bibfnamefont{A.}~\bibnamefont{Tumasyan}} \bibnamefont{et~al.}
  (\bibinfo{collaboration}{CMS}), \bibinfo{journal}{Phys. Rev. D}
  \textbf{\bibinfo{volume}{105}}, \bibinfo{pages}{112007}
  (\bibinfo{year}{2022}), \eprint{2202.08676}.

\bibitem[{\citenamefont{Aad et~al.}(2025)}]{ATLAS:2024mrr}
\bibinfo{author}{\bibfnamefont{G.}~\bibnamefont{Aad}} \bibnamefont{et~al.}
  (\bibinfo{collaboration}{ATLAS}), \bibinfo{journal}{JHEP}
  \textbf{\bibinfo{volume}{05}}, \bibinfo{pages}{075} (\bibinfo{year}{2025}),
  \eprint{2411.07143}.

\bibitem[{\citenamefont{Aad et~al.}(2023)}]{ATLAS:2023sbu}
\bibinfo{author}{\bibfnamefont{G.}~\bibnamefont{Aad}} \bibnamefont{et~al.}
  (\bibinfo{collaboration}{ATLAS}), \bibinfo{journal}{JHEP}
  \textbf{\bibinfo{volume}{07}}, \bibinfo{pages}{118} (\bibinfo{year}{2023}),
  \eprint{2303.05441}.

\bibitem[{\citenamefont{Bahrami et~al.}(2017)\citenamefont{Bahrami, Frank,
  Ghosh, Ghosh, and Saha}}]{Bahrami:2016has}
\bibinfo{author}{\bibfnamefont{S.}~\bibnamefont{Bahrami}},
  \bibinfo{author}{\bibfnamefont{M.}~\bibnamefont{Frank}},
  \bibinfo{author}{\bibfnamefont{D.~K.} \bibnamefont{Ghosh}},
  \bibinfo{author}{\bibfnamefont{N.}~\bibnamefont{Ghosh}}, \bibnamefont{and}
  \bibinfo{author}{\bibfnamefont{I.}~\bibnamefont{Saha}},
  \bibinfo{journal}{Phys. Rev. D} \textbf{\bibinfo{volume}{95}},
  \bibinfo{pages}{095024} (\bibinfo{year}{2017}), \eprint{1612.06334}.

\bibitem[{\citenamefont{Mahmoud et~al.}(2025)\citenamefont{Mahmoud, Kawamura,
  Abdallah, Hussein, and Elgammal}}]{Mahmoud:2024sby}
\bibinfo{author}{\bibfnamefont{Y.}~\bibnamefont{Mahmoud}},
  \bibinfo{author}{\bibfnamefont{J.}~\bibnamefont{Kawamura}},
  \bibinfo{author}{\bibfnamefont{H.}~\bibnamefont{Abdallah}},
  \bibinfo{author}{\bibfnamefont{M.~T.} \bibnamefont{Hussein}},
  \bibnamefont{and} \bibinfo{author}{\bibfnamefont{S.}~\bibnamefont{Elgammal}},
  \bibinfo{journal}{JHEP} \textbf{\bibinfo{volume}{03}}, \bibinfo{pages}{001}
  (\bibinfo{year}{2025}), \eprint{2411.08143}.

\bibitem[{\citenamefont{Yue et~al.}(2025{\natexlab{a}})\citenamefont{Yue, Wang,
  Sun, and Li}}]{Yue:2024ftz}
\bibinfo{author}{\bibfnamefont{C.-X.} \bibnamefont{Yue}},
  \bibinfo{author}{\bibfnamefont{Y.-Q.} \bibnamefont{Wang}},
  \bibinfo{author}{\bibfnamefont{X.-C.} \bibnamefont{Sun}}, \bibnamefont{and}
  \bibinfo{author}{\bibfnamefont{X.-Y.} \bibnamefont{Li}}, \bibinfo{journal}{J.
  Phys. G} \textbf{\bibinfo{volume}{52}}, \bibinfo{pages}{025003}
  (\bibinfo{year}{2025}{\natexlab{a}}), \eprint{2412.07125}.

\bibitem[{\citenamefont{Yue et~al.}(2024)\citenamefont{Yue, Wang, Wang, Wang,
  and Li}}]{Yue:2024sds}
\bibinfo{author}{\bibfnamefont{C.-X.} \bibnamefont{Yue}},
  \bibinfo{author}{\bibfnamefont{Y.-Q.} \bibnamefont{Wang}},
  \bibinfo{author}{\bibfnamefont{H.}~\bibnamefont{Wang}},
  \bibinfo{author}{\bibfnamefont{Y.-H.} \bibnamefont{Wang}}, \bibnamefont{and}
  \bibinfo{author}{\bibfnamefont{S.}~\bibnamefont{Li}}, \bibinfo{journal}{Nucl.
  Phys. B} \textbf{\bibinfo{volume}{1000}}, \bibinfo{pages}{116482}
  (\bibinfo{year}{2024}), \eprint{2402.02072}.

\bibitem[{\citenamefont{Cao et~al.}(2024)\citenamefont{Cao, Guo, Liu, Luo, and
  Wang}}]{Cao:2023smj}
\bibinfo{author}{\bibfnamefont{Q.-H.} \bibnamefont{Cao}},
  \bibinfo{author}{\bibfnamefont{J.}~\bibnamefont{Guo}},
  \bibinfo{author}{\bibfnamefont{J.}~\bibnamefont{Liu}},
  \bibinfo{author}{\bibfnamefont{Y.}~\bibnamefont{Luo}}, \bibnamefont{and}
  \bibinfo{author}{\bibfnamefont{X.-P.} \bibnamefont{Wang}},
  \bibinfo{journal}{Phys. Rev. D} \textbf{\bibinfo{volume}{110}},
  \bibinfo{pages}{015029} (\bibinfo{year}{2024}), \eprint{2311.12934}.

\bibitem[{\citenamefont{Liu and Moretti}(2026)}]{Liu:2025ori}
\bibinfo{author}{\bibfnamefont{Y.-B.} \bibnamefont{Liu}} \bibnamefont{and}
  \bibinfo{author}{\bibfnamefont{S.}~\bibnamefont{Moretti}},
  \bibinfo{journal}{Phys. Rev. D} \textbf{\bibinfo{volume}{113}},
  \bibinfo{pages}{055034} (\bibinfo{year}{2026}), \eprint{2512.22490}.

\bibitem[{\citenamefont{Yue et~al.}(2025{\natexlab{b}})\citenamefont{Yue, Wang,
  and Li}}]{Yue:2025xvk}
\bibinfo{author}{\bibfnamefont{C.-X.} \bibnamefont{Yue}},
  \bibinfo{author}{\bibfnamefont{M.-S.-Y.} \bibnamefont{Wang}},
  \bibnamefont{and} \bibinfo{author}{\bibfnamefont{X.-Y.} \bibnamefont{Li}},
  \bibinfo{journal}{Nucl. Phys. B} \textbf{\bibinfo{volume}{1020}},
  \bibinfo{pages}{117143} (\bibinfo{year}{2025}{\natexlab{b}}).

\bibitem[{\citenamefont{Bhattacharya et~al.}(2022)\citenamefont{Bhattacharya,
  Jahedi, and Wudka}}]{Bhattacharya:2021ltd}
\bibinfo{author}{\bibfnamefont{S.}~\bibnamefont{Bhattacharya}},
  \bibinfo{author}{\bibfnamefont{S.}~\bibnamefont{Jahedi}}, \bibnamefont{and}
  \bibinfo{author}{\bibfnamefont{J.}~\bibnamefont{Wudka}},
  \bibinfo{journal}{JHEP} \textbf{\bibinfo{volume}{05}}, \bibinfo{pages}{009}
  (\bibinfo{year}{2022}), \eprint{2106.02846}.

\bibitem[{\citenamefont{Graham et~al.}(2010)\citenamefont{Graham, Ismail,
  Rajendran, and Saraswat}}]{Graham:2009gy}
\bibinfo{author}{\bibfnamefont{P.~W.} \bibnamefont{Graham}},
  \bibinfo{author}{\bibfnamefont{A.}~\bibnamefont{Ismail}},
  \bibinfo{author}{\bibfnamefont{S.}~\bibnamefont{Rajendran}},
  \bibnamefont{and} \bibinfo{author}{\bibfnamefont{P.}~\bibnamefont{Saraswat}},
  \bibinfo{journal}{Phys. Rev. D} \textbf{\bibinfo{volume}{81}},
  \bibinfo{pages}{055016} (\bibinfo{year}{2010}), \eprint{0910.3020}.

\bibitem[{\citenamefont{Bernreuther and Dobrescu}(2023)}]{Bernreuther:2023uxh}
\bibinfo{author}{\bibfnamefont{E.}~\bibnamefont{Bernreuther}} \bibnamefont{and}
  \bibinfo{author}{\bibfnamefont{B.~A.} \bibnamefont{Dobrescu}},
  \bibinfo{journal}{JHEP} \textbf{\bibinfo{volume}{07}}, \bibinfo{pages}{079}
  (\bibinfo{year}{2023}), \eprint{2304.08509}.

\bibitem[{\citenamefont{Endo et~al.}(2012)\citenamefont{Endo, Hamaguchi,
  Iwamoto, and Yokozaki}}]{Endo:2011xq}
\bibinfo{author}{\bibfnamefont{M.}~\bibnamefont{Endo}},
  \bibinfo{author}{\bibfnamefont{K.}~\bibnamefont{Hamaguchi}},
  \bibinfo{author}{\bibfnamefont{S.}~\bibnamefont{Iwamoto}}, \bibnamefont{and}
  \bibinfo{author}{\bibfnamefont{N.}~\bibnamefont{Yokozaki}},
  \bibinfo{journal}{Phys. Rev. D} \textbf{\bibinfo{volume}{85}},
  \bibinfo{pages}{095012} (\bibinfo{year}{2012}), \eprint{1112.5653}.

\bibitem[{\citenamefont{Ellis et~al.}(2014)\citenamefont{Ellis, Godbole,
  Gopalakrishna, and Wells}}]{Ellis:2014dza}
\bibinfo{author}{\bibfnamefont{S.~A.~R.} \bibnamefont{Ellis}},
  \bibinfo{author}{\bibfnamefont{R.~M.} \bibnamefont{Godbole}},
  \bibinfo{author}{\bibfnamefont{S.}~\bibnamefont{Gopalakrishna}},
  \bibnamefont{and} \bibinfo{author}{\bibfnamefont{J.~D.} \bibnamefont{Wells}},
  \bibinfo{journal}{JHEP} \textbf{\bibinfo{volume}{09}}, \bibinfo{pages}{130}
  (\bibinfo{year}{2014}), \eprint{1404.4398}.

\bibitem[{\citenamefont{Kumar and Martin}(2015)}]{Kumar:2015tna}
\bibinfo{author}{\bibfnamefont{N.}~\bibnamefont{Kumar}} \bibnamefont{and}
  \bibinfo{author}{\bibfnamefont{S.~P.} \bibnamefont{Martin}},
  \bibinfo{journal}{Phys. Rev. D} \textbf{\bibinfo{volume}{92}},
  \bibinfo{pages}{115018} (\bibinfo{year}{2015}), \eprint{1510.03456}.

\bibitem[{\citenamefont{Xu et~al.}(2018)\citenamefont{Xu, Zhang, Li, and
  Li}}]{Xu:2018pnq}
\bibinfo{author}{\bibfnamefont{F.-Z.} \bibnamefont{Xu}},
  \bibinfo{author}{\bibfnamefont{W.}~\bibnamefont{Zhang}},
  \bibinfo{author}{\bibfnamefont{J.}~\bibnamefont{Li}}, \bibnamefont{and}
  \bibinfo{author}{\bibfnamefont{T.}~\bibnamefont{Li}}, \bibinfo{journal}{Phys.
  Rev. D} \textbf{\bibinfo{volume}{98}}, \bibinfo{pages}{115033}
  (\bibinfo{year}{2018}), \eprint{1809.01472}.

\bibitem[{\citenamefont{Dermisek et~al.}(2016)\citenamefont{Dermisek, Lunghi,
  and Shin}}]{Dermisek:2015oja}
\bibinfo{author}{\bibfnamefont{R.}~\bibnamefont{Dermisek}},
  \bibinfo{author}{\bibfnamefont{E.}~\bibnamefont{Lunghi}}, \bibnamefont{and}
  \bibinfo{author}{\bibfnamefont{S.}~\bibnamefont{Shin}},
  \bibinfo{journal}{JHEP} \textbf{\bibinfo{volume}{02}}, \bibinfo{pages}{119}
  (\bibinfo{year}{2016}), \eprint{1509.04292}.

\bibitem[{\citenamefont{Dermisek et~al.}(2014)\citenamefont{Dermisek, Hall,
  Lunghi, and Shin}}]{Dermisek:2014qca}
\bibinfo{author}{\bibfnamefont{R.}~\bibnamefont{Dermisek}},
  \bibinfo{author}{\bibfnamefont{J.~P.} \bibnamefont{Hall}},
  \bibinfo{author}{\bibfnamefont{E.}~\bibnamefont{Lunghi}}, \bibnamefont{and}
  \bibinfo{author}{\bibfnamefont{S.}~\bibnamefont{Shin}},
  \bibinfo{journal}{JHEP} \textbf{\bibinfo{volume}{12}}, \bibinfo{pages}{013}
  (\bibinfo{year}{2014}), \eprint{1408.3123}.

\bibitem[{\citenamefont{Dermisek and Raval}(2013)}]{Dermisek:2013gta}
\bibinfo{author}{\bibfnamefont{R.}~\bibnamefont{Dermisek}} \bibnamefont{and}
  \bibinfo{author}{\bibfnamefont{A.}~\bibnamefont{Raval}},
  \bibinfo{journal}{Phys. Rev. D} \textbf{\bibinfo{volume}{88}},
  \bibinfo{pages}{013017} (\bibinfo{year}{2013}), \eprint{1305.3522}.

\bibitem[{\citenamefont{Freitas et~al.}(2021)\citenamefont{Freitas,
  Gon{\c{c}}alves, Morais, and Pasechnik}}]{Freitas:2020ttd}
\bibinfo{author}{\bibfnamefont{F.~F.} \bibnamefont{Freitas}},
  \bibinfo{author}{\bibfnamefont{J.}~\bibnamefont{Gon{\c{c}}alves}},
  \bibinfo{author}{\bibfnamefont{A.~P.} \bibnamefont{Morais}},
  \bibnamefont{and}
  \bibinfo{author}{\bibfnamefont{R.}~\bibnamefont{Pasechnik}},
  \bibinfo{journal}{JHEP} \textbf{\bibinfo{volume}{01}}, \bibinfo{pages}{076}
  (\bibinfo{year}{2021}), \eprint{2010.01307}.

\bibitem[{\citenamefont{Frampton et~al.}(2000)\citenamefont{Frampton, Hung, and
  Sher}}]{Frampton:1999xi}
\bibinfo{author}{\bibfnamefont{P.~H.} \bibnamefont{Frampton}},
  \bibinfo{author}{\bibfnamefont{P.~Q.} \bibnamefont{Hung}}, \bibnamefont{and}
  \bibinfo{author}{\bibfnamefont{M.}~\bibnamefont{Sher}},
  \bibinfo{journal}{Phys. Rept.} \textbf{\bibinfo{volume}{330}},
  \bibinfo{pages}{263} (\bibinfo{year}{2000}), \eprint{hep-ph/9903387}.

\bibitem[{\citenamefont{Falkowski et~al.}(2014)\citenamefont{Falkowski, Straub,
  and Vicente}}]{Falkowski:2013jya}
\bibinfo{author}{\bibfnamefont{A.}~\bibnamefont{Falkowski}},
  \bibinfo{author}{\bibfnamefont{D.~M.} \bibnamefont{Straub}},
  \bibnamefont{and} \bibinfo{author}{\bibfnamefont{A.}~\bibnamefont{Vicente}},
  \bibinfo{journal}{JHEP} \textbf{\bibinfo{volume}{05}}, \bibinfo{pages}{092}
  (\bibinfo{year}{2014}), \eprint{1312.5329}.

\bibitem[{\citenamefont{Guedes and Santiago}(2022)}]{Guedes:2021oqx}
\bibinfo{author}{\bibfnamefont{G.}~\bibnamefont{Guedes}} \bibnamefont{and}
  \bibinfo{author}{\bibfnamefont{J.}~\bibnamefont{Santiago}},
  \bibinfo{journal}{JHEP} \textbf{\bibinfo{volume}{01}}, \bibinfo{pages}{111}
  (\bibinfo{year}{2022}), \eprint{2107.03429}.

\bibitem[{\citenamefont{Bhattiprolu and Martin}(2019)}]{Bhattiprolu:2019vdu}
\bibinfo{author}{\bibfnamefont{P.~N.} \bibnamefont{Bhattiprolu}}
  \bibnamefont{and} \bibinfo{author}{\bibfnamefont{S.~P.}
  \bibnamefont{Martin}}, \bibinfo{journal}{Phys. Rev. D}
  \textbf{\bibinfo{volume}{100}}, \bibinfo{pages}{015033}
  (\bibinfo{year}{2019}), \eprint{1905.00498}.

\bibitem[{\citenamefont{Sher}(1995)}]{Sher:1995tc}
\bibinfo{author}{\bibfnamefont{M.}~\bibnamefont{Sher}}, \bibinfo{journal}{Phys.
  Rev. D} \textbf{\bibinfo{volume}{52}}, \bibinfo{pages}{3136}
  (\bibinfo{year}{1995}), \eprint{hep-ph/9504257}.

\bibitem[{\citenamefont{Chala et~al.}(2020)\citenamefont{Chala, Koz{\'o}w,
  Ramos, and Titov}}]{Chala:2020odv}
\bibinfo{author}{\bibfnamefont{M.}~\bibnamefont{Chala}},
  \bibinfo{author}{\bibfnamefont{P.}~\bibnamefont{Koz{\'o}w}},
  \bibinfo{author}{\bibfnamefont{M.}~\bibnamefont{Ramos}}, \bibnamefont{and}
  \bibinfo{author}{\bibfnamefont{A.}~\bibnamefont{Titov}},
  \bibinfo{journal}{Phys. Lett. B} \textbf{\bibinfo{volume}{809}},
  \bibinfo{pages}{135752} (\bibinfo{year}{2020}), \eprint{2005.09655}.

\bibitem[{\citenamefont{Guo et~al.}(2023)\citenamefont{Guo, Gao, Mao, and
  Li}}]{Guo:2023jkz}
\bibinfo{author}{\bibfnamefont{Q.}~\bibnamefont{Guo}},
  \bibinfo{author}{\bibfnamefont{L.}~\bibnamefont{Gao}},
  \bibinfo{author}{\bibfnamefont{Y.}~\bibnamefont{Mao}}, \bibnamefont{and}
  \bibinfo{author}{\bibfnamefont{Q.}~\bibnamefont{Li}}, \bibinfo{journal}{Chin.
  Phys. C} \textbf{\bibinfo{volume}{47}}, \bibinfo{pages}{103106}
  (\bibinfo{year}{2023}), \eprint{2304.01885}.

\bibitem[{\citenamefont{Morais et~al.}(2023)\citenamefont{Morais, Onofre,
  Freitas, Gon{\c{c}}alves, Pasechnik, and Santos}}]{Morais:2021ead}
\bibinfo{author}{\bibfnamefont{A.~P.} \bibnamefont{Morais}},
  \bibinfo{author}{\bibfnamefont{A.}~\bibnamefont{Onofre}},
  \bibinfo{author}{\bibfnamefont{F.~F.} \bibnamefont{Freitas}},
  \bibinfo{author}{\bibfnamefont{J.}~\bibnamefont{Gon{\c{c}}alves}},
  \bibinfo{author}{\bibfnamefont{R.}~\bibnamefont{Pasechnik}},
  \bibnamefont{and} \bibinfo{author}{\bibfnamefont{R.}~\bibnamefont{Santos}},
  \bibinfo{journal}{Eur. Phys. J. C} \textbf{\bibinfo{volume}{83}},
  \bibinfo{pages}{232} (\bibinfo{year}{2023}), \eprint{2108.03926}.

\bibitem[{\citenamefont{Ghosh et~al.}(2024)\citenamefont{Ghosh, Rai, and
  Samui}}]{Ghosh:2023xbj}
\bibinfo{author}{\bibfnamefont{N.}~\bibnamefont{Ghosh}},
  \bibinfo{author}{\bibfnamefont{S.~K.} \bibnamefont{Rai}}, \bibnamefont{and}
  \bibinfo{author}{\bibfnamefont{T.}~\bibnamefont{Samui}},
  \bibinfo{journal}{Nucl. Phys. B} \textbf{\bibinfo{volume}{1004}},
  \bibinfo{pages}{116564} (\bibinfo{year}{2024}), \eprint{2309.07583}.

\bibitem[{\citenamefont{Greco et~al.}(2016)\citenamefont{Greco, Han, and
  Liu}}]{Greco:2016izi}
\bibinfo{author}{\bibfnamefont{M.}~\bibnamefont{Greco}},
  \bibinfo{author}{\bibfnamefont{T.}~\bibnamefont{Han}}, \bibnamefont{and}
  \bibinfo{author}{\bibfnamefont{Z.}~\bibnamefont{Liu}},
  \bibinfo{journal}{Phys. Lett. B} \textbf{\bibinfo{volume}{763}},
  \bibinfo{pages}{409} (\bibinfo{year}{2016}), \eprint{1607.03210}.

\bibitem[{\citenamefont{Delahaye et~al.}(2019)\citenamefont{Delahaye, Diemoz,
  Long, Mansouli{\'e}, Pastrone, Rivkin, Schulte, Skrinsky, and
  Wulzer}}]{Delahaye:2019omf}
\bibinfo{author}{\bibfnamefont{J.~P.} \bibnamefont{Delahaye}},
  \bibinfo{author}{\bibfnamefont{M.}~\bibnamefont{Diemoz}},
  \bibinfo{author}{\bibfnamefont{K.}~\bibnamefont{Long}},
  \bibinfo{author}{\bibfnamefont{B.}~\bibnamefont{Mansouli{\'e}}},
  \bibinfo{author}{\bibfnamefont{N.}~\bibnamefont{Pastrone}},
  \bibinfo{author}{\bibfnamefont{L.}~\bibnamefont{Rivkin}},
  \bibinfo{author}{\bibfnamefont{D.}~\bibnamefont{Schulte}},
  \bibinfo{author}{\bibfnamefont{A.}~\bibnamefont{Skrinsky}}, \bibnamefont{and}
  \bibinfo{author}{\bibfnamefont{A.}~\bibnamefont{Wulzer}}
  (\bibinfo{year}{2019}), \eprint{1901.06150}.

\bibitem[{\citenamefont{Long et~al.}(2021)\citenamefont{Long, Lucchesi, Palmer,
  Pastrone, Schulte, and Shiltsev}}]{Long:2020wfp}
\bibinfo{author}{\bibfnamefont{K.}~\bibnamefont{Long}},
  \bibinfo{author}{\bibfnamefont{D.}~\bibnamefont{Lucchesi}},
  \bibinfo{author}{\bibfnamefont{M.}~\bibnamefont{Palmer}},
  \bibinfo{author}{\bibfnamefont{N.}~\bibnamefont{Pastrone}},
  \bibinfo{author}{\bibfnamefont{D.}~\bibnamefont{Schulte}}, \bibnamefont{and}
  \bibinfo{author}{\bibfnamefont{V.}~\bibnamefont{Shiltsev}},
  \bibinfo{journal}{Nature Phys.} \textbf{\bibinfo{volume}{17}},
  \bibinfo{pages}{289} (\bibinfo{year}{2021}), \eprint{2007.15684}.

\bibitem[{\citenamefont{de~Blas et~al.}(2022)}]{MuonCollider:2022xlm}
\bibinfo{author}{\bibfnamefont{J.}~\bibnamefont{de~Blas}} \bibnamefont{et~al.}
  (\bibinfo{collaboration}{Muon Collider}) (\bibinfo{year}{2022}),
  \eprint{2203.07261}.

\bibitem[{\citenamefont{Accettura et~al.}(2023)}]{Accettura:2023ked}
\bibinfo{author}{\bibfnamefont{C.}~\bibnamefont{Accettura}}
  \bibnamefont{et~al.}, \bibinfo{journal}{Eur. Phys. J. C}
  \textbf{\bibinfo{volume}{83}}, \bibinfo{pages}{864} (\bibinfo{year}{2023}),
  \bibinfo{note}{[Erratum: Eur.Phys.J.C 84, 36 (2024)]}, \eprint{2303.08533}.

\bibitem[{\citenamefont{Accettura
  et~al.}(2024)}]{InternationalMuonCollider:2024jyv}
\bibinfo{author}{\bibfnamefont{C.}~\bibnamefont{Accettura}}
  \bibnamefont{et~al.} (\bibinfo{collaboration}{International Muon Collider}),
  \bibinfo{journal}{CERN Yellow Rep. Monogr.}
  \textbf{\bibinfo{volume}{2/2024}}, \bibinfo{pages}{176}
  (\bibinfo{year}{2024}), \eprint{2407.12450}.

\bibitem[{\citenamefont{Black et~al.}(2024)}]{Black:2022cth}
\bibinfo{author}{\bibfnamefont{K.~M.} \bibnamefont{Black}}
  \bibnamefont{et~al.}, \bibinfo{journal}{JINST} \textbf{\bibinfo{volume}{19}},
  \bibinfo{pages}{T02015} (\bibinfo{year}{2024}), \eprint{2209.01318}.

\bibitem[{\citenamefont{Han et~al.}(2021{\natexlab{a}})\citenamefont{Han, Ma,
  and Xie}}]{Han:2020uid}
\bibinfo{author}{\bibfnamefont{T.}~\bibnamefont{Han}},
  \bibinfo{author}{\bibfnamefont{Y.}~\bibnamefont{Ma}}, \bibnamefont{and}
  \bibinfo{author}{\bibfnamefont{K.}~\bibnamefont{Xie}},
  \bibinfo{journal}{Phys. Rev. D} \textbf{\bibinfo{volume}{103}},
  \bibinfo{pages}{L031301} (\bibinfo{year}{2021}{\natexlab{a}}),
  \eprint{2007.14300}.

\bibitem[{\citenamefont{Costantini et~al.}(2020)\citenamefont{Costantini,
  De~Lillo, Maltoni, Mantani, Mattelaer, Ruiz, and Zhao}}]{Costantini:2020stv}
\bibinfo{author}{\bibfnamefont{A.}~\bibnamefont{Costantini}},
  \bibinfo{author}{\bibfnamefont{F.}~\bibnamefont{De~Lillo}},
  \bibinfo{author}{\bibfnamefont{F.}~\bibnamefont{Maltoni}},
  \bibinfo{author}{\bibfnamefont{L.}~\bibnamefont{Mantani}},
  \bibinfo{author}{\bibfnamefont{O.}~\bibnamefont{Mattelaer}},
  \bibinfo{author}{\bibfnamefont{R.}~\bibnamefont{Ruiz}}, \bibnamefont{and}
  \bibinfo{author}{\bibfnamefont{X.}~\bibnamefont{Zhao}},
  \bibinfo{journal}{JHEP} \textbf{\bibinfo{volume}{09}}, \bibinfo{pages}{080}
  (\bibinfo{year}{2020}), \eprint{2005.10289}.

\bibitem[{\citenamefont{Huang et~al.}(2022)\citenamefont{Huang, Jana, Queiroz,
  and Rodejohann}}]{Huang:2021biu}
\bibinfo{author}{\bibfnamefont{G.-y.} \bibnamefont{Huang}},
  \bibinfo{author}{\bibfnamefont{S.}~\bibnamefont{Jana}},
  \bibinfo{author}{\bibfnamefont{F.~S.} \bibnamefont{Queiroz}},
  \bibnamefont{and}
  \bibinfo{author}{\bibfnamefont{W.}~\bibnamefont{Rodejohann}},
  \bibinfo{journal}{Phys. Rev. D} \textbf{\bibinfo{volume}{105}},
  \bibinfo{pages}{015013} (\bibinfo{year}{2022}), \eprint{2103.01617}.

\bibitem[{\citenamefont{Bandyopadhyay et~al.}(2024)\citenamefont{Bandyopadhyay,
  Parashar, Sen, and Song}}]{Bandyopadhyay:2024plc}
\bibinfo{author}{\bibfnamefont{P.}~\bibnamefont{Bandyopadhyay}},
  \bibinfo{author}{\bibfnamefont{S.}~\bibnamefont{Parashar}},
  \bibinfo{author}{\bibfnamefont{C.}~\bibnamefont{Sen}}, \bibnamefont{and}
  \bibinfo{author}{\bibfnamefont{J.}~\bibnamefont{Song}},
  \bibinfo{journal}{JHEP} \textbf{\bibinfo{volume}{07}}, \bibinfo{pages}{253}
  (\bibinfo{year}{2024}), \eprint{2401.02697}.

\bibitem[{\citenamefont{Yang et~al.}(2021)\citenamefont{Yang, Qian, Guan, Li,
  Meng, Xiao, Lu, and Li}}]{Yang:2021zak}
\bibinfo{author}{\bibfnamefont{T.}~\bibnamefont{Yang}},
  \bibinfo{author}{\bibfnamefont{S.}~\bibnamefont{Qian}},
  \bibinfo{author}{\bibfnamefont{Z.}~\bibnamefont{Guan}},
  \bibinfo{author}{\bibfnamefont{C.}~\bibnamefont{Li}},
  \bibinfo{author}{\bibfnamefont{F.}~\bibnamefont{Meng}},
  \bibinfo{author}{\bibfnamefont{J.}~\bibnamefont{Xiao}},
  \bibinfo{author}{\bibfnamefont{M.}~\bibnamefont{Lu}}, \bibnamefont{and}
  \bibinfo{author}{\bibfnamefont{Q.}~\bibnamefont{Li}}, \bibinfo{journal}{Phys.
  Rev. D} \textbf{\bibinfo{volume}{104}}, \bibinfo{pages}{093003}
  (\bibinfo{year}{2021}), \eprint{2107.13581}.

\bibitem[{\citenamefont{Han et~al.}(2021{\natexlab{b}})\citenamefont{Han, Liu,
  Low, and Wang}}]{Han:2020pif}
\bibinfo{author}{\bibfnamefont{T.}~\bibnamefont{Han}},
  \bibinfo{author}{\bibfnamefont{D.}~\bibnamefont{Liu}},
  \bibinfo{author}{\bibfnamefont{I.}~\bibnamefont{Low}}, \bibnamefont{and}
  \bibinfo{author}{\bibfnamefont{X.}~\bibnamefont{Wang}},
  \bibinfo{journal}{Phys. Rev. D} \textbf{\bibinfo{volume}{103}},
  \bibinfo{pages}{013002} (\bibinfo{year}{2021}{\natexlab{b}}),
  \eprint{2008.12204}.

\bibitem[{\citenamefont{Cingiloglu and Frank}(2025)}]{Cingiloglu:2024vdh}
\bibinfo{author}{\bibfnamefont{K.~Y.} \bibnamefont{Cingiloglu}}
  \bibnamefont{and} \bibinfo{author}{\bibfnamefont{M.}~\bibnamefont{Frank}},
  \bibinfo{journal}{Phys. Rev. D} \textbf{\bibinfo{volume}{111}},
  \bibinfo{pages}{016025} (\bibinfo{year}{2025}), \eprint{2408.10898}.

\bibitem[{\citenamefont{Adhikary et~al.}(2024)\citenamefont{Adhikary,
  Olechowski, Rosiek, and Ryczkowski}}]{Adhikary:2024esf}
\bibinfo{author}{\bibfnamefont{A.}~\bibnamefont{Adhikary}},
  \bibinfo{author}{\bibfnamefont{M.}~\bibnamefont{Olechowski}},
  \bibinfo{author}{\bibfnamefont{J.}~\bibnamefont{Rosiek}}, \bibnamefont{and}
  \bibinfo{author}{\bibfnamefont{M.}~\bibnamefont{Ryczkowski}},
  \bibinfo{journal}{Phys. Rev. D} \textbf{\bibinfo{volume}{110}},
  \bibinfo{pages}{075029} (\bibinfo{year}{2024}), \eprint{2406.16050}.

\bibitem[{\citenamefont{Cynolter and Lendvai}(2008)}]{Cynolter:2008ea}
\bibinfo{author}{\bibfnamefont{G.}~\bibnamefont{Cynolter}} \bibnamefont{and}
  \bibinfo{author}{\bibfnamefont{E.}~\bibnamefont{Lendvai}},
  \bibinfo{journal}{Eur. Phys. J. C} \textbf{\bibinfo{volume}{58}},
  \bibinfo{pages}{463} (\bibinfo{year}{2008}), \eprint{0804.4080}.

\bibitem[{\citenamefont{Lavoura and Silva}(1993)}]{Lavoura:1992np}
\bibinfo{author}{\bibfnamefont{L.}~\bibnamefont{Lavoura}} \bibnamefont{and}
  \bibinfo{author}{\bibfnamefont{J.~P.} \bibnamefont{Silva}},
  \bibinfo{journal}{Phys. Rev. D} \textbf{\bibinfo{volume}{47}},
  \bibinfo{pages}{2046} (\bibinfo{year}{1993}).

\bibitem[{\citenamefont{Alloul et~al.}(2014)\citenamefont{Alloul, Christensen,
  Degrande, Duhr, and Fuks}}]{Alloul:2013bka}
\bibinfo{author}{\bibfnamefont{A.}~\bibnamefont{Alloul}},
  \bibinfo{author}{\bibfnamefont{N.~D.} \bibnamefont{Christensen}},
  \bibinfo{author}{\bibfnamefont{C.}~\bibnamefont{Degrande}},
  \bibinfo{author}{\bibfnamefont{C.}~\bibnamefont{Duhr}}, \bibnamefont{and}
  \bibinfo{author}{\bibfnamefont{B.}~\bibnamefont{Fuks}},
  \bibinfo{journal}{Comput. Phys. Commun.} \textbf{\bibinfo{volume}{185}},
  \bibinfo{pages}{2250} (\bibinfo{year}{2014}), \eprint{1310.1921}.

\bibitem[{\citenamefont{Degrande et~al.}(2012)\citenamefont{Degrande, Duhr,
  Fuks, Grellscheid, Mattelaer, and Reiter}}]{Degrande:2011ua}
\bibinfo{author}{\bibfnamefont{C.}~\bibnamefont{Degrande}},
  \bibinfo{author}{\bibfnamefont{C.}~\bibnamefont{Duhr}},
  \bibinfo{author}{\bibfnamefont{B.}~\bibnamefont{Fuks}},
  \bibinfo{author}{\bibfnamefont{D.}~\bibnamefont{Grellscheid}},
  \bibinfo{author}{\bibfnamefont{O.}~\bibnamefont{Mattelaer}},
  \bibnamefont{and} \bibinfo{author}{\bibfnamefont{T.}~\bibnamefont{Reiter}},
  \bibinfo{journal}{Comput. Phys. Commun.} \textbf{\bibinfo{volume}{183}},
  \bibinfo{pages}{1201} (\bibinfo{year}{2012}), \eprint{1108.2040}.

\bibitem[{\citenamefont{Alwall et~al.}(2014)\citenamefont{Alwall, Frederix,
  Frixione, Hirschi, Maltoni, Mattelaer, Shao, Stelzer, Torrielli, and
  Zaro}}]{Alwall:2014hca}
\bibinfo{author}{\bibfnamefont{J.}~\bibnamefont{Alwall}},
  \bibinfo{author}{\bibfnamefont{R.}~\bibnamefont{Frederix}},
  \bibinfo{author}{\bibfnamefont{S.}~\bibnamefont{Frixione}},
  \bibinfo{author}{\bibfnamefont{V.}~\bibnamefont{Hirschi}},
  \bibinfo{author}{\bibfnamefont{F.}~\bibnamefont{Maltoni}},
  \bibinfo{author}{\bibfnamefont{O.}~\bibnamefont{Mattelaer}},
  \bibinfo{author}{\bibfnamefont{H.~S.} \bibnamefont{Shao}},
  \bibinfo{author}{\bibfnamefont{T.}~\bibnamefont{Stelzer}},
  \bibinfo{author}{\bibfnamefont{P.}~\bibnamefont{Torrielli}},
  \bibnamefont{and} \bibinfo{author}{\bibfnamefont{M.}~\bibnamefont{Zaro}},
  \bibinfo{journal}{JHEP} \textbf{\bibinfo{volume}{07}}, \bibinfo{pages}{079}
  (\bibinfo{year}{2014}), \eprint{1405.0301}.

\bibitem[{\citenamefont{Sj{\"o}strand et~al.}(2015)\citenamefont{Sj{\"o}strand,
  Ask, Christiansen, Corke, Desai, Ilten, Mrenna, Prestel, Rasmussen, and
  Skands}}]{Sjostrand:2014zea}
\bibinfo{author}{\bibfnamefont{T.}~\bibnamefont{Sj{\"o}strand}},
  \bibinfo{author}{\bibfnamefont{S.}~\bibnamefont{Ask}},
  \bibinfo{author}{\bibfnamefont{J.~R.} \bibnamefont{Christiansen}},
  \bibinfo{author}{\bibfnamefont{R.}~\bibnamefont{Corke}},
  \bibinfo{author}{\bibfnamefont{N.}~\bibnamefont{Desai}},
  \bibinfo{author}{\bibfnamefont{P.}~\bibnamefont{Ilten}},
  \bibinfo{author}{\bibfnamefont{S.}~\bibnamefont{Mrenna}},
  \bibinfo{author}{\bibfnamefont{S.}~\bibnamefont{Prestel}},
  \bibinfo{author}{\bibfnamefont{C.~O.} \bibnamefont{Rasmussen}},
  \bibnamefont{and} \bibinfo{author}{\bibfnamefont{P.~Z.}
  \bibnamefont{Skands}}, \bibinfo{journal}{Comput. Phys. Commun.}
  \textbf{\bibinfo{volume}{191}}, \bibinfo{pages}{159} (\bibinfo{year}{2015}),
  \eprint{1410.3012}.

\bibitem[{\citenamefont{de~Favereau et~al.}(2014)\citenamefont{de~Favereau,
  Delaere, Demin, Giammanco, Lema\^\i{}tre, Mertens, and
  Selvaggi}}]{deFavereau:2013fsa}
\bibinfo{author}{\bibfnamefont{J.}~\bibnamefont{de~Favereau}},
  \bibinfo{author}{\bibfnamefont{C.}~\bibnamefont{Delaere}},
  \bibinfo{author}{\bibfnamefont{P.}~\bibnamefont{Demin}},
  \bibinfo{author}{\bibfnamefont{A.}~\bibnamefont{Giammanco}},
  \bibinfo{author}{\bibfnamefont{V.}~\bibnamefont{Lema\^\i{}tre}},
  \bibinfo{author}{\bibfnamefont{A.}~\bibnamefont{Mertens}}, \bibnamefont{and}
  \bibinfo{author}{\bibfnamefont{M.}~\bibnamefont{Selvaggi}}
  (\bibinfo{collaboration}{DELPHES 3}), \bibinfo{journal}{JHEP}
  \textbf{\bibinfo{volume}{02}}, \bibinfo{pages}{057} (\bibinfo{year}{2014}),
  \eprint{1307.6346}.

\bibitem[{Muo()}]{MuonColliderDelphesCard}
\bibinfo{howpublished}{\url{https://github.com/delphes/delphes/blob/master/cards/delphes_card_MuonColliderDet.tcl}}.

\bibitem[{\citenamefont{Boronat et~al.}(2015)\citenamefont{Boronat, Fuster,
  Garcia, Ros, and Vos}}]{Boronat:2014hva}
\bibinfo{author}{\bibfnamefont{M.}~\bibnamefont{Boronat}},
  \bibinfo{author}{\bibfnamefont{J.}~\bibnamefont{Fuster}},
  \bibinfo{author}{\bibfnamefont{I.}~\bibnamefont{Garcia}},
  \bibinfo{author}{\bibfnamefont{E.}~\bibnamefont{Ros}}, \bibnamefont{and}
  \bibinfo{author}{\bibfnamefont{M.}~\bibnamefont{Vos}},
  \bibinfo{journal}{Phys. Lett. B} \textbf{\bibinfo{volume}{750}},
  \bibinfo{pages}{95} (\bibinfo{year}{2015}), \eprint{1404.4294}.

\bibitem[{\citenamefont{Boronat et~al.}(2018)\citenamefont{Boronat, Fuster,
  Garcia, Roloff, Simoniello, and Vos}}]{Boronat:2016tgd}
\bibinfo{author}{\bibfnamefont{M.}~\bibnamefont{Boronat}},
  \bibinfo{author}{\bibfnamefont{J.}~\bibnamefont{Fuster}},
  \bibinfo{author}{\bibfnamefont{I.}~\bibnamefont{Garcia}},
  \bibinfo{author}{\bibfnamefont{P.}~\bibnamefont{Roloff}},
  \bibinfo{author}{\bibfnamefont{R.}~\bibnamefont{Simoniello}},
  \bibnamefont{and} \bibinfo{author}{\bibfnamefont{M.}~\bibnamefont{Vos}},
  \bibinfo{journal}{Eur. Phys. J. C} \textbf{\bibinfo{volume}{78}},
  \bibinfo{pages}{144} (\bibinfo{year}{2018}), \eprint{1607.05039}.

\bibitem[{\citenamefont{Conte et~al.}(2013)\citenamefont{Conte, Fuks, and
  Serret}}]{Conte:2012fm}
\bibinfo{author}{\bibfnamefont{E.}~\bibnamefont{Conte}},
  \bibinfo{author}{\bibfnamefont{B.}~\bibnamefont{Fuks}}, \bibnamefont{and}
  \bibinfo{author}{\bibfnamefont{G.}~\bibnamefont{Serret}},
  \bibinfo{journal}{Comput. Phys. Commun.} \textbf{\bibinfo{volume}{184}},
  \bibinfo{pages}{222} (\bibinfo{year}{2013}), \eprint{1206.1599}.

\bibitem[{\citenamefont{Conte et~al.}(2014)\citenamefont{Conte, Dumont, Fuks,
  and Wymant}}]{Conte:2014zja}
\bibinfo{author}{\bibfnamefont{E.}~\bibnamefont{Conte}},
  \bibinfo{author}{\bibfnamefont{B.}~\bibnamefont{Dumont}},
  \bibinfo{author}{\bibfnamefont{B.}~\bibnamefont{Fuks}}, \bibnamefont{and}
  \bibinfo{author}{\bibfnamefont{C.}~\bibnamefont{Wymant}},
  \bibinfo{journal}{Eur. Phys. J. C} \textbf{\bibinfo{volume}{74}},
  \bibinfo{pages}{3103} (\bibinfo{year}{2014}), \eprint{1405.3982}.

\bibitem[{\citenamefont{Conte and Fuks}(2018)}]{Conte:2018vmg}
\bibinfo{author}{\bibfnamefont{E.}~\bibnamefont{Conte}} \bibnamefont{and}
  \bibinfo{author}{\bibfnamefont{B.}~\bibnamefont{Fuks}},
  \bibinfo{journal}{Int. J. Mod. Phys. A} \textbf{\bibinfo{volume}{33}},
  \bibinfo{pages}{1830027} (\bibinfo{year}{2018}), \eprint{1808.00480}.

\bibitem[{\citenamefont{Cowan et~al.}(2011)\citenamefont{Cowan, Cranmer, Gross,
  and Vitells}}]{Cowan:2010js}
\bibinfo{author}{\bibfnamefont{G.}~\bibnamefont{Cowan}},
  \bibinfo{author}{\bibfnamefont{K.}~\bibnamefont{Cranmer}},
  \bibinfo{author}{\bibfnamefont{E.}~\bibnamefont{Gross}}, \bibnamefont{and}
  \bibinfo{author}{\bibfnamefont{O.}~\bibnamefont{Vitells}},
  \bibinfo{journal}{Eur. Phys. J. C} \textbf{\bibinfo{volume}{71}},
  \bibinfo{pages}{1554} (\bibinfo{year}{2011}), \bibinfo{note}{[Erratum:
  Eur.Phys.J.C 73, 2501 (2013)]}, \eprint{1007.1727}.

\bibitem[{\citenamefont{Basso}(2022)}]{Basso:2021dwp}
\bibinfo{author}{\bibfnamefont{M.~J.} \bibnamefont{Basso}},
  \bibinfo{journal}{J. Phys. G} \textbf{\bibinfo{volume}{49}},
  \bibinfo{pages}{025001} (\bibinfo{year}{2022}), \eprint{2102.04275}.

\end{thebibliography}
\end{document}